\documentclass[useAMS,usenatbib]{mn2e}
\usepackage{psfig}

\title[Quasars]{Optical identification of XMM sources in the CFHTLS}
\author[C. S. Stalin et al.]{C. S. Stalin$^{1}$\thanks{E-mail: stalin@iiap.res.in}, Patrick Petitjean$^{2}$, R. Srianand$^{3}$,
A. J. Fox$^{4}$, F. Coppolani$^{2}$, A. Schwope$^{5}$  \\ 
$^{1}$Indian Institute of Astrophysics, Koramangala, Bangalore 560\,034, India\\
$^{2}$Institut d'Astrophysique de Paris, CNRS - Universit\'e Pierre et Marie Curie, 98bis bd Arago, 75014 Paris, France\\
$^{3}$Inter University Center for Astronomy and Astrophysics, Post Bag 4, Ganesh Khind, Pune 411\,007, India \\
$^{4}$European Southern Observatory, Alonso de C\'{o}rdova 3107, Casilla 19001, Vitacura, Santiago 19, Chile\\
$^{5}$Astrophysikalishes Institut Potsdam, An der Sternwarte 16, D-14482 Potsdam, Germany\\}

\begin{document}

\date{}

\pagerange{\pageref{firstpage}--\pageref{lastpage}} \pubyear{2008}

\maketitle

\label{firstpage}

\begin{abstract}
We present optical spectroscopic identifications of X-ray sources in  
$\sim$3 square degrees of the XMM-Large Scale Structure survey (XMM-LSS), also 
covered by the Canada France Hawaii Telescope Legacy Survey (CFHTLS),
obtained with the AAOmega instrument at the Anglo Australian Telescope. 
In a flux limited sample of 829 point like sources in
the optical band with $g^{\prime}$~$\le$~22 mag and  
the 0.5 $-$ 2 keV flux ($f_{\rm 0.5 - 2 keV}$) $>$ 
1 $\times$ 10$^{-15}$~erg~cm$^{-2}$ s$^{-1}$, 
we observed 695 objects and obtained reliable spectroscopic 
identification for 489 sources, $\sim59\%$ of the overall sample. 
We therefore increase the number of identifications in this field by
a factor close to five. Galactic stellar 
sources represent about 15\% of the total (74/489). About 55\% (267/489) are 
broad-line Active Galactic Nuclei (AGNs) spanning redshifts between 
0.15 and 3.87 with a median value of 
1.68. The optical-to-X-ray spectral index ($\alpha_{\rm ox}$) of the  
broad-line AGNs is 1.47 $\pm$ 0.03, 
typical of optically-selected Type I quasars and is found 
to correlate with the rest frame X-ray and optical monochromatic 
luminosities at 2 keV and 2500~{\AA} respectively. 
%We do not observe any 
%evolution of this 
%index with redshift for these broad-line AGNs. This is consistent with earlier studies available in the 
%literature. 
Consistent with previous studies, we find $\alpha_{ox}$ not to be
correlated with {\it z}. In addition, 32 and 116 X-ray sources are, respectively absorption and 
emission-line galaxies at $z$~$<$~0.76.
%Absorption line galaxies are all at $z$ 
%$<$ 0.8, while the emission-line galaxies are at $z$ between 0.02 and 0.76. 
From a line ratio diagnostic diagram it is found that in about 50\% of 
these emission line galaxies, the emission lines are powered 
significantly by the AGN. 
Thirty of the XMM sources are detected 
at one or more radio frequencies.  In addition, 24 sources have ambiguous 
identification: in 8 cases, two XMM sources have a single optical
source within 6$^{\prime\prime}$ of each of them,  
whereas, 2 and 14 XMM sources have, respectively, 3 and 2 possible 
optical sources within 6$^{\prime\prime}$ of each of them. 
%We report the 
%spectra of all the optical sources within 6$^{\prime\prime}$  of these 16 XMM
%sources. 
Spectra of multiple possible counterparts were obtained in such 
ambiguous cases.
\end{abstract}


\section{Introduction}
Sky surveys play a key role in astronomy as they provide the basic data 
%which allow one 
for the characterization of different populations of astronomical 
objects and their evolution over cosmic time. 
%There are currently a number of on-going sky surveys over a wide wavelength
%range such as in optical, IR, X-ray etc. of which 
We take advantage here of two surveys of the same field: 
 (i) the XMM-Large Scale Structure Survey (XMM-LSS; Pierre et al. 2004) to survey the X-ray sky to a relatively
low flux limit of 1$\times$10$^{-15}$ erg s$^{-1}$ cm$^{-2}$ in the 0.5$-$2 keV band and 
(ii) the Wide part of Canada France Hawaii Telescope Legacy 
Survey (CFHTLS; Cuillandre \& Bertin 2006) to observe 
limited portions of the sky to faint magnitude limits in five optical bands.

The vast majority of the AGNs known today were discovered during optical surveys
like the bright Palomar Quasar Survey (PG; Schmidt \& Green 1983), the Large Bright Quasar Survey 
(LBQS; Hewett et al. 1995), the Hamburg-ESO Quasar Survey (HES; Reimers, Koehler \& Wisotzki 1996)
the Two Degree Field Quasar Survey (2dF; Croom et al. 2001) and 
the Sloan Digital Sky Survey (SDSS; Schneider et al. 2007). Such optical colour selection
based on wide photometric bands can miss the obscured AGNs. The 
SDSS survey 
has identified many type 2 AGNs based on the presence of narrow emission
lines in their spectra without broad lines, and with emission line ratios 
typical of AGN (Kauffmann et al. 2003; Zakamska et al. 2003). Alternatively, 
X-ray surveys include AGNs without narrow emission lines or those at high 
redshift where emission line selection is not possible. This is why the identification 
of X-ray sources down to low optical magnitudes
is an important step towards elucidating AGN activity.

There are a number of existing surveys in various energy bands in the X-ray wavelength
range such as the EMSS (Einstein Extended Medium Sensitivity Survey; Maccacaro
et al. 1982; Stocke et al. 1991), the ROSAT Bright Survey (Schwope et al. 2000)
and the ASCA survey (Ueda et al. 2003), but they all are of relatively poor 
sensitivity. There are 
however deep surveys with ROSAT such as the UDS (ROSAT Ultra Deep Survey; 
Lehmann et al. 2001) and the RDS (ROSAT Deep Survey; Hasinger et al. 1998). 
They reach an X-ray flux limit of $\sim$10$^{-15}$ erg cm$^{-2}$s$^{-1}$, however
over very small regions in the sky. 
With the {\it Chandra} and XMM {\it Newton} X-ray telescopes
in operation, X-ray surveys have been significantly boosted. Surveys 
carried out
using XMM-Newton and {\it Chandra} include the SXDS (Subaru/XMM-Newton
Deep Survey; Ueda et al. 2008), LALA (Large Area Lyman Alpha Survey; 
Wang et al. 2004), CLASXS (Chandra Large Area Synoptic Survey, Yang 
et al. 2004), ChaMP(Chandra Multiwavelength Project; Kim et al. 2004), 
CDF-S (Chandra Deep Field-South; Luo et al. 2008; Szokoly et al. 2004), 
CDF-N (Chandra Deep
Field-North; Barger et al. 2003), Bootes Survey (Brand et al. 2006), 
HELLA2XMM (Cocchia et al. 2007) and the XMM-COSMOS 
survey (Brusa et al. 2007; Cappelluti et al. 2009; Trump et al. 2009).
Most of these X-ray surveys are either deep/ultra-deep pencil beam
surveys or shallow large area surveys. 
%Large area moderately deep X-ray
%surveys will be a good complementarity to the narrow area deep surveys and
%large area shallow surveys as they can obtain a more complete coverage of 
%extragalactic sources in the luminosity-redshift plane.
%One such large area
%survey conducted with XMM {\it Newton} is a medium depth survey in the CFHTLS region, the XMM-LSS.

The XMM-LSS survey is a medium depth survey conducted with XMM Newton
in the CFHTLS region.
It is designed to provide a well defined statistical sample of
X-ray selected galaxy clusters out to a redshift of unity, over a large
area suitable for cosmological studies (Pierre et al. 2004).
Apart from finding new galaxy clusters, XMM-LSS will also provide 
X-ray point-like sources down to low flux levels, of which AGNs might represent
$\sim$95\% (Pierre et al. 2007) as they are
known to be strong X-ray emitters (Fabbiano et al. 1992) in
comparison to normal galaxies. These AGNs are thought to be at the origin of 
most of the X-ray background (Giacconi et al. 2002; Alexander et al. 2003).
Recently, about 80\% of the background has been resolved in the 2$-$10 keV 
energy range by deep {\it Chandra} and XMM-Newton observations (e.g. Worsley et al. 2005; Hickox
\& Markevitch 2006; Carrera et al. 2007). 
Optical identification of such XMM sources is of great importance to characterize the AGN 
population as a function of redshift and address specific issues
related to the importance of the AGN phenomenon in galaxy formation and evolution
of the intergalactic medium.
There is an overlap between the XMM-LSS and surveys at different wavelengths:
the VIMOS VLT Deep Survey (VVDS; Le F\`evre et al. 2004) and the CFHTLS in the optical, 
the UKIRT Infrared Deep Sky Survey (UKIDSS; Dye et al. 2006; Lawrence et al. 2007) 
in the near IR, the {\it Spitzer} Wide Area InfraRed Extragalactic Legacy Survey (SWIRE; 
Lonsdale et al. 2003), radio observations from the VLA at 1.4 GHz (Bondi et al. 2003) and 
at 325 and 74 MHz (Cohen et al. 2003), 610 MHz observations from GMRT (Bondi et al. 2007) 
and UV observations with GALEX (Arnouts et al. 2005; Schiminovich et al. 2005).
In this work, we use version 1 of the XMM-LSS X-ray catalogue (Pierre et al. 2007). Sources were entered in the catalogue 
if the likelihood of detection in either of the survey bands is greater than 15, and if the observed flux is
larger than a 0.5 $-$ 2keV flux limit of 1 $\times$ 10$^{-15}$ erg cm$^{-2}$ s$^{-1}$. 
%This version 1 of the X-ray catalogue  also lists optical counterparts from CFHTLS to those X-ray sources. 

%X-ray surveys enable one to understand the 
%X-ray background, the main component of which is again thought to be made
%of AGNs (Giacconi et al. 2002; Alexander et al. 2003).
%However, not all AGNs selected from surveys  
%in the optical, IR and/or radio have counterparts in X-rays (Kollatschmy et al. 2008). 

% and therefore not efficient in finding the actual number of AGN. 
%On the other hand, X-ray, IR and radio surveys will be efficient in having a complete
%census of AGN as they are less affected by obscuration effects. However, 
%X-ray and IR surveys have either smaller areas and less AGN or 
%wider area but shallower than optical surveys. 

%they will enable one to find, apart from the Type I AGN sources, the obscured fraction 
%of Type II AGN.Thus, from an optical identification of  X-ray sources detected in XMM-LSS, 
%the whole AGN population can be effectively selected and studied. However this required dedicated 
%optical spectroscopic follow up observations of these X-ray sources to clearly identify them.
%Once selected, AGN can be used as probes to study the history of the early universe. They can also be used to 

%address other specific issues such as determination of the nature of the quasar population as 
%a function of redshift, radio-loud/radio-quiet dichotomy, understanding quasar formation and evolution
%through quasar luminosity function, their contribution to X-ray, UV background radiation etc. 
\begin{figure}
\hspace*{-0.5cm}\psfig{file=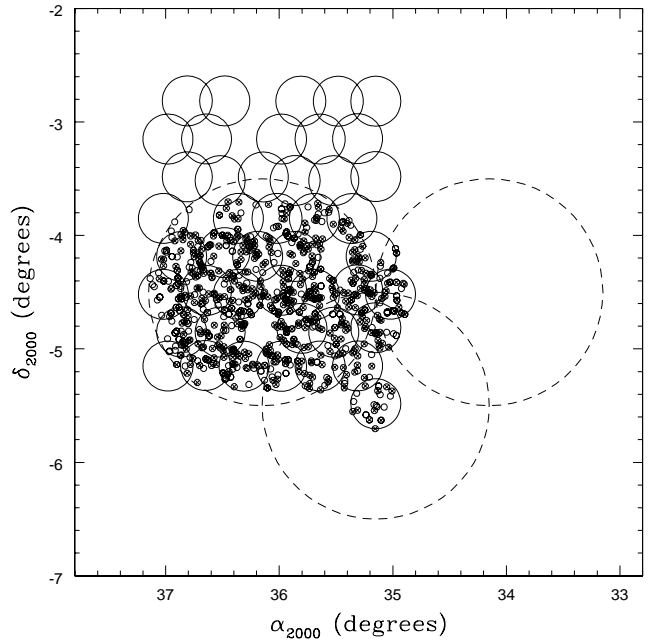,width=9cm,height=9cm}
\caption{Layout of the 45 XMM-LSS pointings (large closed circles). The positions
of our initial sample of 829 sources with $g^{\prime}$~$<$~22 mag and flux in the 0.5$-$2 keV band larger than
1$\times$10$^{-15}$ erg cm$^{-2}$ s$^{-1}$ are marked with open 
small circles. Crosses show the XMM sources for which reliable spectroscopic identifications are 
obtained in this work. The upper half of the XMM-LSS pointings do not 
have corresponding optical coverage in CFHTLS. The larger dashed circles are
the three AAT pointings}
\end{figure}

\begin{figure}
\hspace*{-0.1cm}\psfig{file=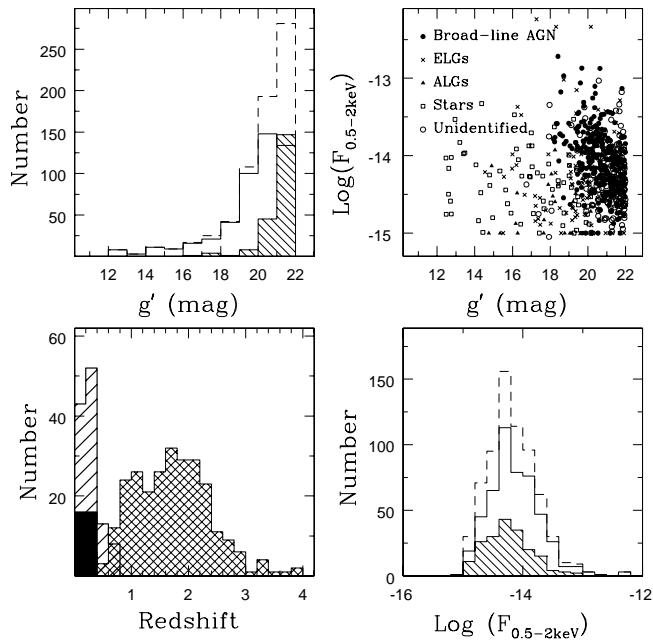,width=9cm,height=9cm}
\caption{{\sl Top left panel}: $g^{\prime}$ magnitude distribution of 
XMM sources observed in this work (dashed histogram), 
sources with certain 
identifications (solid histogram) and un-identified sources (hatched histogram).
{\sl Bottom left panel}: Redshift distribution 
of broad-line AGNs (cross hatched region), emission-line objects (hatched region) 
 and absorption-line galaxies (shaded region). {\sl Top right panel}: plot of 
the X-ray flux at 0.5$-$2keV (erg cm$^{-2}$ s$^{-1}$) band against the 
optical $g^{\prime}$ magnitude
for broad-line AGNs (filled circles), emission line galaxies (crosses),
absorption line galaxies (filled triangles), stars (open squares) and 
un-identified sources
(open circles). {\sl Bottom right panel}: X-ray flux distribution in 
0.5$-$2keV (erg cm$^{-2}$ s$^{-1}$)
band for our initial sample (dashed histogram), the spectroscopically identified
sources (solid histogram) and the un-identified sources (hatched region).}
\end{figure}

\begin{figure}
\hspace*{-0.2cm}\psfig{file=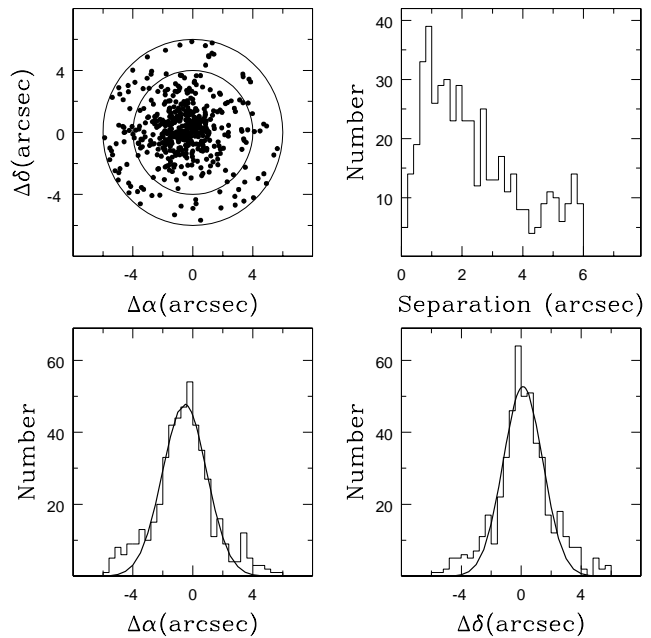,width=9cm,height=9cm}
\caption{{\it Top panels:} Offsets between the XMM and optical (XMM $-$ optical) 
positions (left) and the distribution of the angular separation between the 
XMM and the optical positions (right) of the spectroscopically identified X-ray 
sources. Circles in the top left panel have radius of 4$^{\prime\prime}$ and
6$^{\prime\prime}$ respectively. {\it Bottom panels:} Offset histograms for
$\Delta \alpha$ (left) and $\Delta \delta$ (right). The solid line is 
the gaussian fit to the distribution of offsets. Note the fit is not a 
good representation of the data and an offset of about 0.15$^{\prime\prime}$
is present in the $\alpha$ direction.}
\end{figure}

\begin{figure}
\hspace*{-1.0cm}\psfig{file=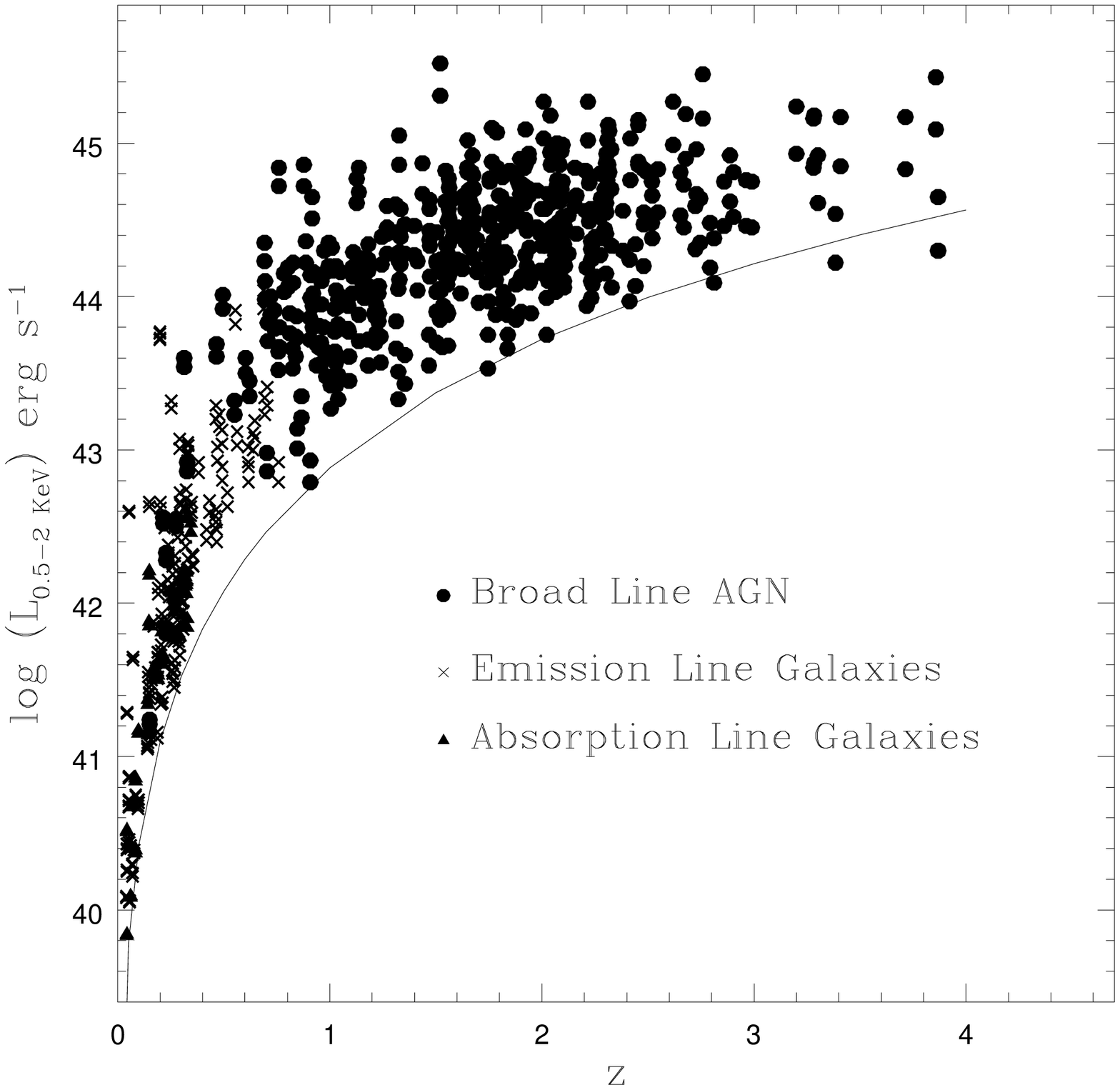,width=9cm,height=9cm}
\caption{The observed 0.5$-$2 keV X-ray luminosity of the XMM sources as a 
function of redshift. Filled circles are for broad-line AGNs,  
crosses are for emission-line galaxies and 
filled triangles are for absorption-line galaxies. The line shows the 
luminosity 
calculated as a function of redshift for a 0.5$-$2 keV
flux limit of 1$\times$10$^{-15}$ erg cm$^{-2}$ s$^{-1}$. }
\end{figure}

In this paper we present optical spectroscopic identification of these
XMM sources in $\sim$3 square degrees of the CFHTLS down to $g<22$ mag. 
This spectroscopic campaign is intermediate between large scale but shallower surveys
like the 2dF (Croom et al. 2001) or SDSS (Schneider et al. 2007) and
deeper but spatially restricted surveys like the  
VVDS (Gavignaud et al. 2006) which has a limiting magnitude of 
I$_{\rm AB}$~$<$~24, the AGN survey in the COSMOS field (Trump et al. 2009) reaching
a magnitude limit of i~$<$~23.5 and optical identification of XMM sources
in the ELIAS-S1 field reaching down to R~$\sim$24.5 (Feruglio et al. 2008).
%over $\sim$0.5 square degree each. 
%of identifying XMM sources with the aim of identifying AGNs 
%is  deeper than the other large area optical surveys such as the 2dF (Croom et al. 2001) and
%the SDSS (Schneider et al. 2007) however fainter than the other known
%fainter AGN spectroscopic campaigns  

Identifications of X-ray sources in the XMM-LSS have been already obtained by
Tajer et al. (2006) and Garcet et al. (2007). The former authors survey
sources detected in 1~square degree with $F_{2-10keV}$ $>$ 10$^{-14}$ erg cm$^{-2}$ s$^{-1}$ 
in the 2$-$10 keV band at a significance $\ge$3$\sigma$ and report 122 unambiguous
identifications. The later authors present identification of 99 sources with
R$\le$22 mag. Here we increase by a factor close to five the number
of identifications in this field. The paper is
organized as follows. We present the selection of targets in Section 2; Section 3 describes
the spectroscopic observations; identifications are discussed in Section 4; 
Section 5 gives the discussion of results and conclusions are summarized in the final section.
Throughout this paper we adopt a cosmology with
$H_{\rm o}$ = 70 km s$^{-1}$ Mpc$^{-1}$, $\Omega_{\rm M}$ = 0.27 and $\Omega_{\Lambda}$ = 0.73.

\section{Optical and X-ray surveys}
%\subsection{The Canada France Hawaii Telescope Legacy Survey(CFHTLS)
The basic data sets used to select the targets for this survey are 
from the XMM-LSS (Pierre et al. 2004) and the
wide synoptic component (W1) of 
CFHTLS.\footnote{http://www.cfht.hawaai.edu/Science/CFHTLS/}
%and the XMM Large Scale Structure Survey (XMM-LSS; Pierre et al. 2004). 
%
\subsection{CFHTLS} 
The CFHTLS is an ambitious 
imaging program that has been carried out at the 3.6m Canada-France Hawaii Telescope 
using the wide field prime focus MegaPrime equipped with MEGACAM, a 36 CCD
mosaic camera each of them with 2048$\times$4612 pixels$^2$. The pixel scale
is 0.185$^{\prime\prime}$, thus giving a total field of view of 0.96 $\times$ 0.94 deg$^2$. 
CFHTLS consists of a deep survey in 4 fields (D1, D2, D3 and D4) each covering 
about 1$\times$1 square degree and a shallower survey on four wide fields (W1, W2, W3, W4) 
each covering 7$\times$7 square degrees in u$^{\ast}$, g$^{\prime}$, r$^{\prime}$,
i$^{\prime}$ and z$^{\prime}$ filters. Final limiting magnitudes in the wide fields should be
of the same order in the five bands and typically $i\sim24.5$. The internal accuracy
of the astrometric solution (band to band) is better than one pixel rms over the 
entire MEGACAM field, whereas the external astrometric solution is around 
0.25$^{\prime\prime}$ rms (Schultheiss et al. 2006).
%CFHTLS can thus be a very valuable source to look for optical counterparts to other surveys (X-ray, IR) overlapping 
%with its survey region.

\begin{figure}
\hspace*{-0.8cm}\psfig{file=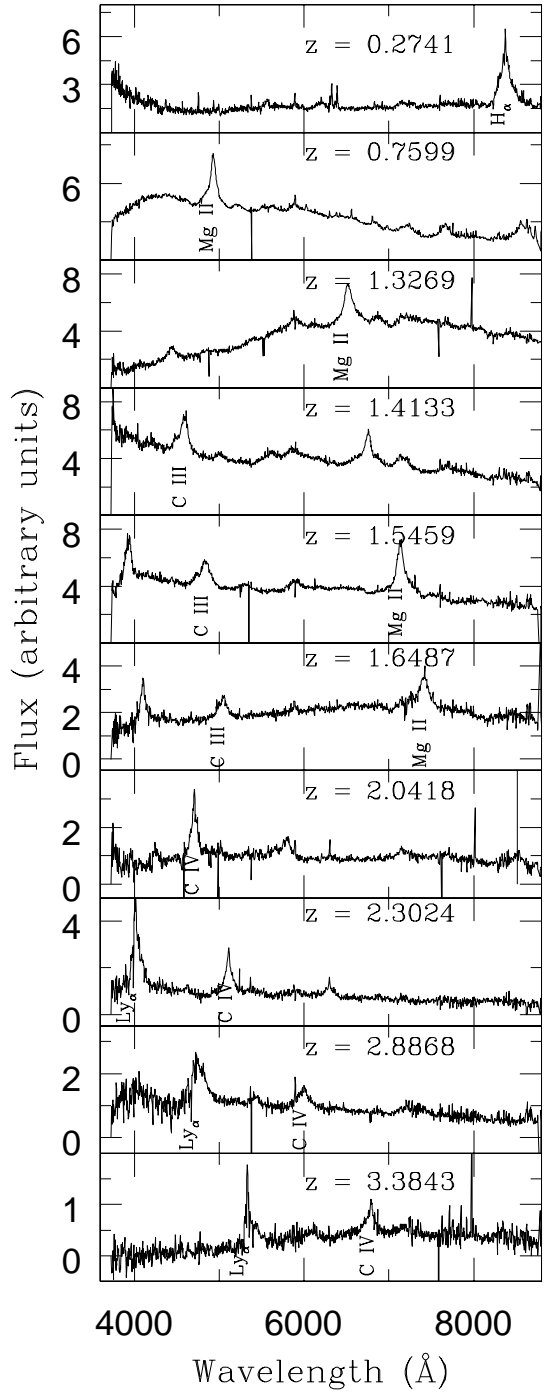}
\caption{Examples of broad-line AGN optical spectra. In each panels we
mark the locations of prominent emission lines and the corresponding 
redshift of the source.}
\end{figure}

\begin{figure}
\hspace*{-0.8cm}\psfig{file=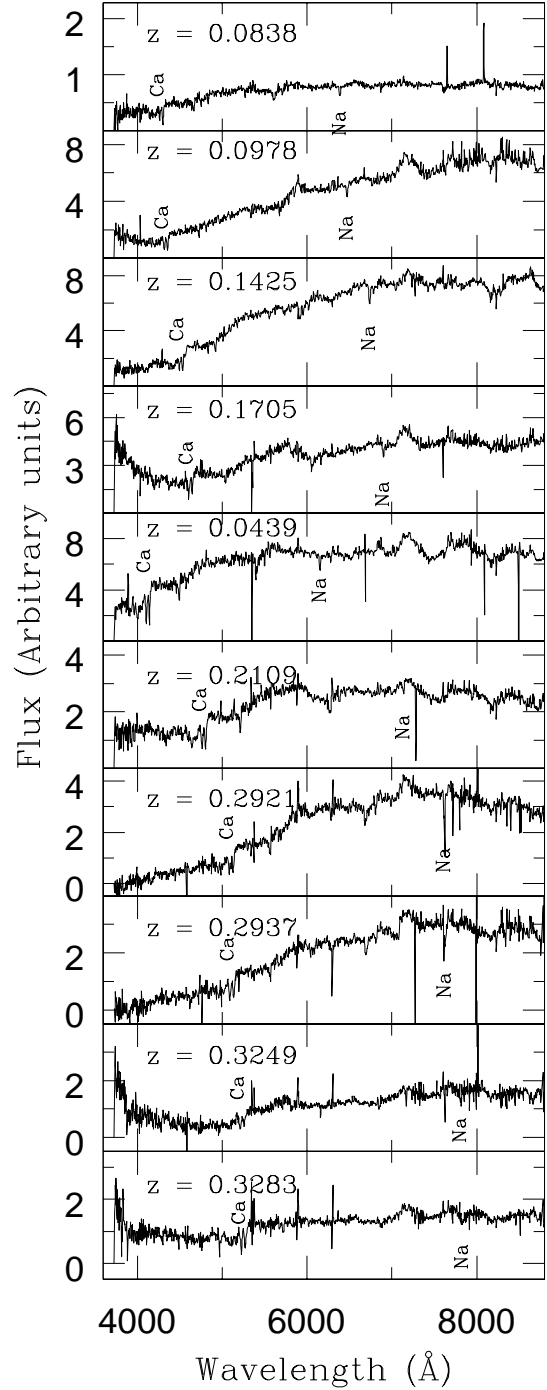}
\caption{Examples of optical spectra of absorption-line galaxies. 
Redshifts are indicated 
on each panels. The  Na absorption line is marked and 
the region where Ca H \& K lines  are found is indicated as Ca.}
\end{figure}

\begin{figure}
\hspace*{-0.8cm}\psfig{file=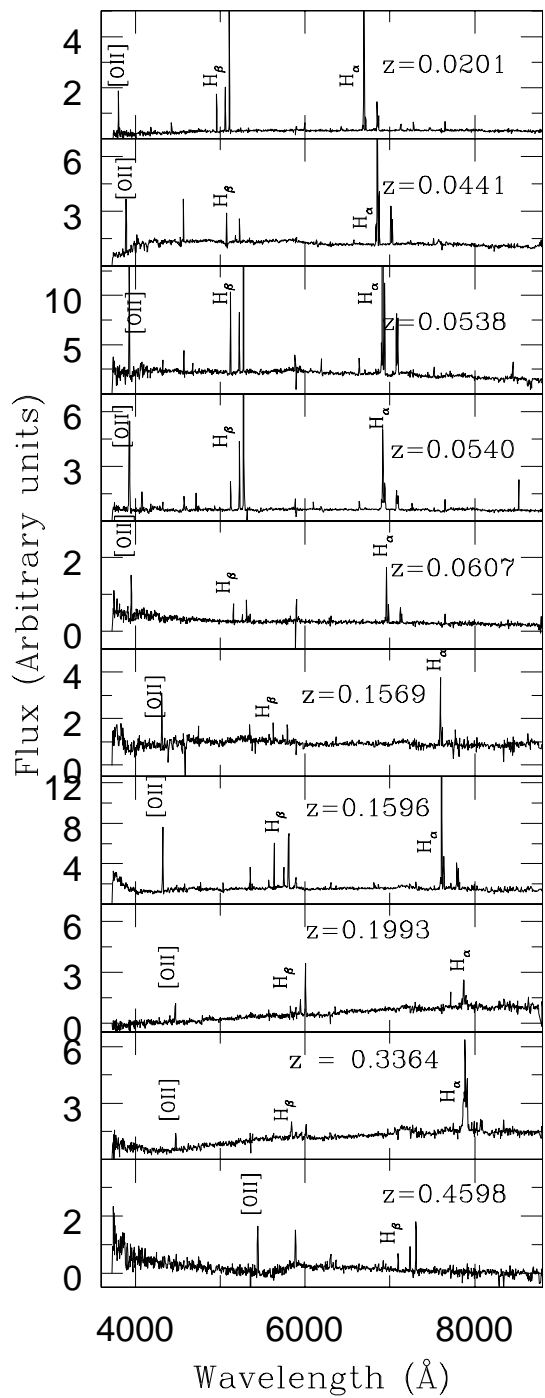}
\caption{Examples of optical spectra of emission-line galaxies. 
The redshift of each galaxy is given in the corresponding panel.  
The prominent emission lines are marked}and 
\end{figure}

\begin{figure}
\hspace{-1.0cm}\psfig{file=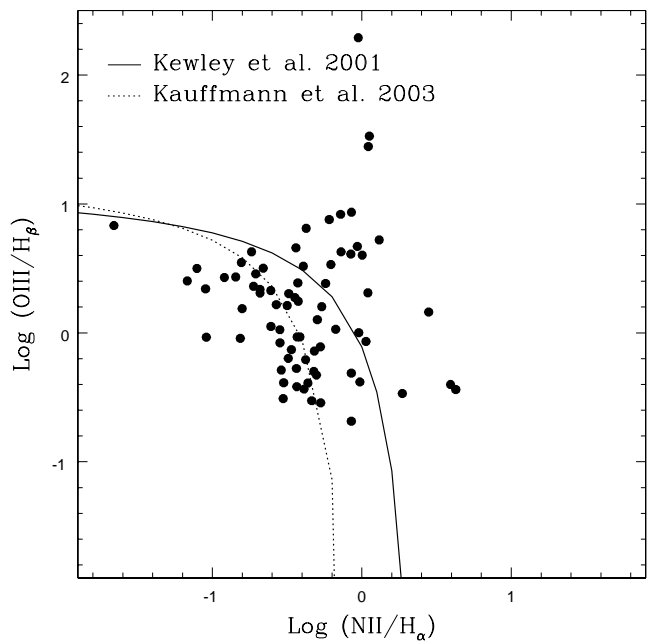,width=9cm,height=9cm}
\caption{The Baldwin-Phillips-Terlevich (BPT) diagnostic  diagram 
for the emission line objects with [OIII], H${\beta}$, [NII] and
H${\alpha}$ detected in their optical spectra. Objects above and below
the lines are powered by AGN and starbursts respectively. The solid
line is that of Kewley et al. (2001) and the dotted line is that
of Kauffmann et al. (2003).}
\end{figure}

\begin{figure}
\hspace*{-0.6cm}\psfig{file=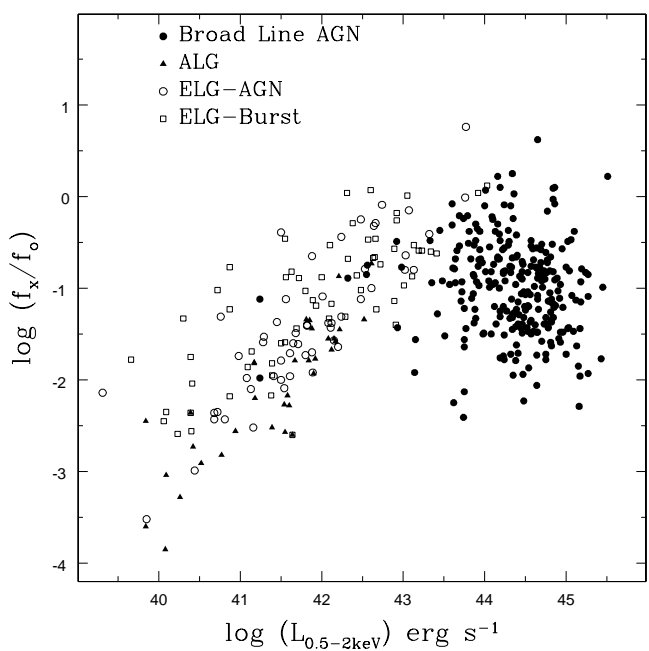,width=9cm,height=9cm}
\caption{X-ray to optical flux ratio ($f_x/f_o$) plotted against the intrinsic 
X-ray luminosity $L_{0.5-2keV}$}
\end{figure}

\begin{figure}
\hspace*{-0.5cm}\psfig{file=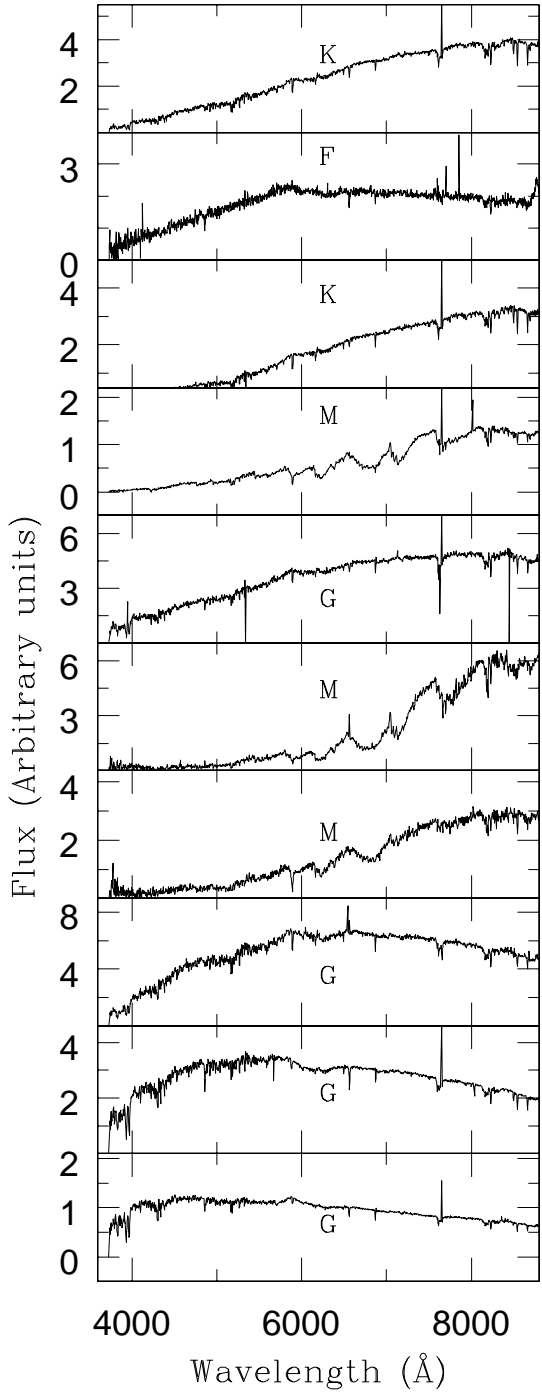}
\caption{Examples of the optical spectra of stars. Their spectral types are given on 
each panel.}
\end{figure}

\begin{figure}
\hspace*{-1.0cm}\psfig{file=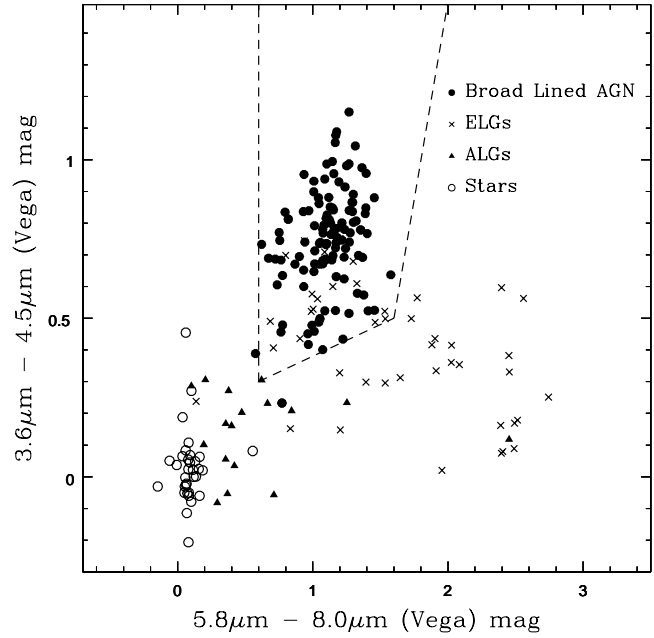,width=9cm,height=9cm}
\caption{Positions of the various kinds of XMM sources in 
the IR colour$-$colour diagram. Here  
filled circles are broad-line AGNs, crosses are emission-line galaxies, 
filled diamonds are absorption-line galaxies and open 
circles are stars. The dotted line is the empirical region of
Stern et al. (2005) used to separate AGN from stars and galaxies.} 
\end{figure}

\begin{figure}
\hspace*{-0.6cm}\psfig{file=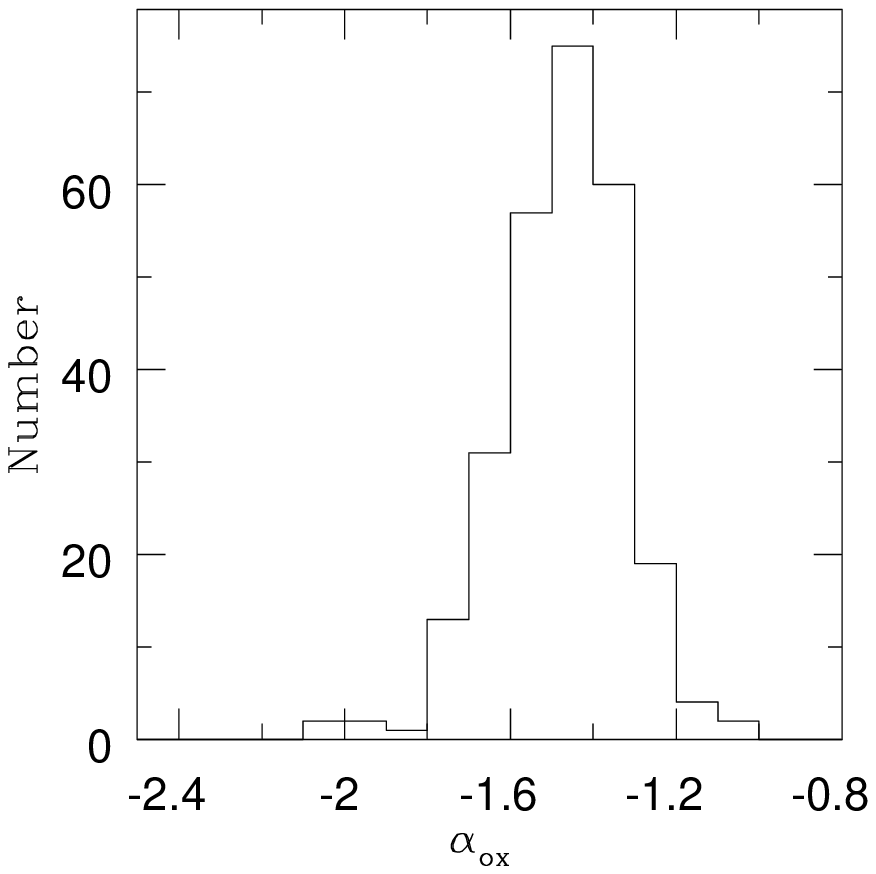,width=9cm,height=9cm}
\caption{Distribution of $\alpha_{\rm ox}$ for the spectroscopically
identified broad-line AGNs in our sample of XMM sources.}
\end{figure}

\subsection{XMM-LSS} 
XMM-LSS is a medium depth large area X-ray survey designed to map large scale structures 
in the Universe (Pierre et al. 2004) and is located around coordinates, RA = 2$^h$ and Dec = $-$4$^{\deg}$ (see Fig.~1). 
Pierre et al. (2007) present the source list obtained from observations of the first 5.5 square degrees 
(pertaining to 45 XMM pointings) in the 0.5$-$2~keV and 2$-$10~keV  bands and 
above a detection likelihood of 15 in either bands. They also provide the list of optical 
objects extracted from the CFHTLS catalogue and located within a radius of 6 arcsec around each X-ray source.  
The layout of the XMM-LSS pointings (big circles) is shown in Fig.~1. We 
extracted from this catalogue 
the list of X-ray sources 
%with a flux larger than 1$\times$10$^{-15}$ erg cm$^{-2}$ s$^{-1}$ and with 
having optical counterparts from CFHTLS brighter than $g^{\prime}$~$=$~22 mag. 
We thus arrived at 829 XMM sources having an optical counterpart 
within 6$^{\prime\prime}$ of the X-ray source and brighter than 
$g^{\prime}$~$=$~22 mag.
The positions of these sources
are marked in Fig.~1 with open small circles. Also marked with crosses 
on these open small circles are the sources 
that are reliably identified in the course of this work.
%for which reliable spectroscopic identifications are known from this work.

%The layout of XMM pointings on the
%CFHTLS region (small circles) along with the region taken
%up for optical spectroscopic followup (large circle) is shown in Fig. ~1.

\section{Observations}
The spectroscopic observations were performed with the AAOmega system (Sharp et
al. 2006) at the 3.9~m Anglo-Australian Telescope (AAT), during the
nights of 25, 26, 27 September 2006 and 11, 12, 13 September 2007. AAT 
observations were focussed on three regions and are shown as three
larger circles in Fig. 2. AAOmega
involves the  multi-object fibre feed from the 2dF fibre positioner system
(Lewis et al. 2002) linked to an efficient and stable
bench-mounted dual-beam spectrograph. The fibre positioner can place
392 fibres of 2 arcsec projected diameter within a 2 degree
field of view. Within each configuration there is a minimum target
separation of 30 arcsec imposed by the physical size of the fibre
buttons (Miszalski et al. 2006). The dual-beam AAOmega
spectrograph, with  both blue and red arms, was used in its default
low resolution configuration. In the blue arm, the 580V volume phase holographic 
(VPH) grating was used, providing wavelength coverage between 3700 to 5800 \AA~ 
sampled at 1~\AA/pixel. In the red arm, the 385R VPH grating was used, providing
wavelength coverage from 5600 to 8800~\AA~ sampled at 1.6~\AA/pixel.
The red and blue segments are  spliced together between 5600 and 5800~\AA~
thus giving continuous wavelength coverage betwen 3700 to 8800~\AA~ at a
resolution of $\sim$1300. Of the 392 fibres, 25 fibres were assigned to random
positions uniformly distributed over the 2 square degrees to sample the
background sky  and 8 fibre positions were assigned to guide stars
within the field. For each plate configuration (each 2 degree field was observed 
in two to three plate configurations), the observations  were broken into  blocks of
upto 2 hours to minimize flux losses. Each block consists
of a quartz-halogen flat field exposure, a composite arc lamp frame for
wavelength calibration and three to four science exposures. Total exposure
time on each target varies from 10 minutes to 2 hours depending on their
brightness. 
The spectra were not flux calibrated. 
%and as such flux calibration 
%is not necessary for spectral identificaion of XMM sources. 

Reduction of the observed spectra  was performed using the AAOmega's data reduction 
pipeline software  DRCONTROL. The two dimensional images were flatfielded, and the 
spectrum was extracted (using a gaussian profile extraction), wavelength calibrated and combined within
DRCONTROL. Redshift estimate was performed using the AUTOZ code (this was kindly provided by Scott Croom).
All spectra, identifications and redshift determinations were checked manually. 
%by several of us.
%visual
%inspection and line ratio diagnostic diagrams (BPT diagrams; Baldwin et al. ).
The reduced
optical spectra of the 489 identified XMM sources are available upon 
request.

\section{Identification and classification of X-ray sources}
From the initial sample of 829 optical counterparts to XMM sources brighter
than $g^{\prime}$ = 22 mag, we were able to obtain spectra
for a total of 695 sources.  
%of a total of 695 optical counterparts of X-ray sources 
%selected as described above were obtained. Of these,
%so as to identify their nature of which 
Of these, 489 spectra ($\sim$70\%) are of sufficient S/N to identify their 
nature. 
Standard classification schemes (e.g. Caccianiga et al. 2008) were used to 
classify these objects from 
the spectral features detected in their spectra.  
The $g^{\prime}$-band magnitude distribution of our observed sample
of 695 sources (dashed histogram),  the 489 spectroscopically 
identified XMM sources (solid histogram) and the 206 unidentified sources (hatched
region) are shown in the top left panel of Fig.~2. The bottom left panel of
Fig.~2 shows the redshift distributions of broad-line AGNs (cross hatched 
region), 
emission line galaxies (sparsely shaded region)  and 
absorption-line galaxies (darkely shaded region). In the top right 
panel of 
Fig.~2 is shown the X-ray flux (0.5$-$2keV) against the optical 
$g^{\prime}$-band magnitude for broad-line AGNs (filled circles), emission
line galaxies (crosses), absorption line galaxies (filled triangles), 
stars (open squares) and unidentified sources (open circles). The bottom 
right panel of Fig.~2 shows the 0.5$-$2keV flux distribution of the 
observed sources (dashed histogram), the spectroscopically identified
sources (solid histogram) and the unidentified sources (hatched region)
respectively.
%of emission and absorption
%line sources, the classification of which is done as detailed below.
% 
The distribution of the angular separation between the X-ray and optical
positions of the 489 sources is shown in Fig. ~3. 
%Also shown in Fig. ~3 
%are the X-ray and optical positional offsets for these 489 sources. In Fig. ~3 we
%also show the histogram of the offsets in $\alpha$ and $\delta$ between
%the optical and XMM positions along with a gaussian fit to the 
%distribution. 
Also shown in this figure are the histograms of the offsets in 
$\alpha$ and $\delta$ ($\alpha_{XMM}$ $-$ $\alpha_{optical}$; $\delta_{XMM}$ $-$ $\delta_{optical}$) 
along with a gaussian fit to the distributions.
%The fitted gaussian function is close to the true distribution
%to offsets $\sim$4$^{\prime\prime}$, with mismatch at separations
%larger than  4$^{\prime\prime}$. 406 sources (83\%) are within the 4$^{\prime\prime}$
%error circle in Fig. ~3. and their identification as optical counterparts
%to XMM sources could be secure. It is thus possible that some of the
%optical counterparts with larger offsets ($>$ 4$^{\prime\prime}$) are 
%likely misidentifications. 
The fitted gaussian function does not reproduce well the real distribution
beyond 4$^{\prime\prime}$. 406 sources (83\%) are within the 4$^{\prime\prime}$
error circle and their identification is probably secure. It is possible 
that some of the optical counterparts with larger offets ($>$ 4$^{\prime\prime}$) are not correctly identified.
The $\Delta \delta$ distribution is found
to be symmetrically distributed around zero, whereas the 
 $\Delta \alpha$ distribution is offset by about 0.15$^{\prime\prime}$.
This shift in $\alpha$ might be due to systematics in the astrometric
corrections applied in the XMM-LSS catalogue (Pierre et al. 2007).
%a shift in the gaussian distribution of $\sim$0.15$^{\prime\prime}$ is 
%noticed.  
The intrinsic X-ray luminosity of 
these sources plotted against their redshift is shown in Fig. ~4 together with the 
curve marking the position of a source with a
0.5$-$2 keV flux limit of 1$\times$10$^{-15}$ erg cm$^{-2}$ s$^{-1}$. 
%It is found that the BLAGN show a broad redshift 
%distribution distribution between $z < xx < xx $ with a median $z$ = xxx.
%ELGs are found in the redshift range $z < xx < xx $ with a median $z$ = xxx.

Among the spectroscopically classified 489 
sources, we find 267 broad-line AGNs (BLAGNs; $\sim$55\% of the identifications), 116 
emission-line galaxies (ELGs; $\sim$24\% of the sample), 32 absorption-line galaxies
(ALGs; $\sim$6\% of the sample), and 74 stars ($\sim$15\% of the sample). 
Among the 116 ELGs, based on Baldwin-Phillips-Terlevich (BPT) diagnostic 
diagram (Baldwin, Phillips, Terlevich 1981), 
%55 are found to be significantly 
%powered by AGN 
emission lines in 55 ELGs are consistent with them being powered
significantly by AGN and 61 by starbursts
(see below). A summary of these numbers is given in Table 1. Details of 
these 489 objects are given in Table 2. For most of the sources
with spectral lines in our spectroscopic sample, we were able to obtain
a reliable estimate of the redshift. This is because for these sources, we were
able to identify two or more lines in their spectra. We assign a quality flag 
(indicative of the reliability of the redshift) Q = 1 for those sources. For 
sources, with only one broad or narrow emission/absorption line in their optical spectra, 
the redshift estimate is not secure as it is degenerate with more than one possible
redshifts. For such sources, we assign a quality flag Q = 2.
%Table 2 gives the list
%of XMM sources which are identified as BLAGN. The list of indentified
%narrow line objects (including NLAGN and ELGs) are given in Table 3. List of 
%absorption-line galaxies and stars are given in Tables 4 and 5 respectively.
For some sources, the spectral classification is unambiguous, however, the
spectra is of poor quality. For these sources, we have assigned a 
quality flag Q=3. For 206 sources, spectral classification was not possible
due to the poor quality of their spectra. Such sources have faint optical 
magnitudes and this is clearly seen in the top left panel in Fig.~2. In
our 489 identified sources, 434, 7 and 48 sources have quality 
flags Q1, Q2 and Q3 respectively.

\begin{table}
 \centering
 \begin{minipage}{70mm}
\caption{Statistics of the spectroscopic identifications of XMM sources. 
Column 1 is the object type, column 2 is the number of unambiguous
identifications and column 3 gives the respective percentage.}
\begin{tabular}{ccc} \hline
Object Type& Total Number & Percentage \\    
           &                 &              \\ \hline
BLAGN      &   267            & 38.4       \\
ELG        &   116            & 16.7        \\
ALG        &   32             &  4.6       \\
Stars      &   74             & 10.6       \\
Unidentified &  206           & 29.7       \\  \hline
Total      &      695         & 100.0      \\ \hline
\end{tabular}
\end{minipage}
\end{table}

\begin{table*}
 \centering
 \begin{minipage}{140mm}
\caption{Properties of the optical identification of the XMM sources. Only 
the first ten entries are shown. The table in its entirety is available 
in the electronic edition. Here column 1 is XMM ID, column 2 is optical $\alpha$, 
column 3 is optical $\delta$, column 4 is the angular separation between the
XMM and optical position, column 5 is $g^{\prime}$ magnitude, 
column 6 is redshift, column 7 is the quality flag of the spectra, 
column 8 is $\alpha_{ox}$ and column 9 is our spectroscopic identification. 
Objects detected by Spitzer are indicated as SWIRE in the last column}
\begin{tabular}{cccccccccc} \hline
   XMM ID              & $\alpha_{2000}$& $\delta_{2000}$ & sep              &   $g^{\prime}$   & z & Q & $\alpha_{ox}$ & ID &  Spitzer \\
                       & (optical)      & (optical)       & ($\prime\prime$) & (mag) &   &   &               &    &   \\ \hline
XLSS J021934.6-044140  & 2:19:34.702  & -4:41:41.022  &     1.16   &    20.84   &     2.10   &   1 &    -1.42  & QSO &    SWIRE   \\
XLSS J021935.6-042543  & 2:19:35.932  & -4:25:46.048  &     5.12   &    20.95   &     2.48   &   1 &    -1.38  & QSO &            \\
XLSS J021941.1-044059  & 2:19:41.160  & -4:41:00.311  &     0.69   &    20.77   &     2.10   &   1 &    -1.30  & QSO &    SWIRE   \\
XLSS J021946.9-043754  & 2:19:47.121  & -4:37:54.632  &     1.89   &    20.06   &     1.62   &   1 &    -1.56  & QSO &    SWIRE   \\
XLSS J021951.2-043417  & 2:19:51.233  & -4:34:16.203  &     0.88   &    20.84   &     1.91   &   1 &    -1.35  & QSO &    SWIRE   \\
XLSS J021952.0-040918  & 2:19:52.149  & -4:09:19.867  &     1.99   &    20.37   &     0.69   &   1 &    -1.21  & QSO &    SWIRE   \\
XLSS J021952.3-042447  & 2:19:52.374  & -4:24:48.675  &     1.84   &    21.70   &     0.56   &   1 &    -1.34  & ELG &            \\
XLSS J021957.2-043952  & 2:19:57.248  & -4:39:52.421  &     0.59   &    18.25   &     2.07   &   1 &    -1.67  & QSO &    SWIRE   \\
XLSS J021958.1-041712  & 2:19:58.133  & -4:17:07.659  &     4.62   &    21.12   &     1.84   &   3 &    -1.62  & QSO &    SWIRE   \\
XLSS J022000.1-041746  & 2:20:00.159  & -4:17:45.391  &     0.86   &    21.81   &     0.92   &   1 &    -1.00  & QSO &    SWIRE   \\ \hline
\end{tabular}
\end{minipage}
\end{table*}

\subsection{Broad line AGNs}
Objects having at least one of the emission lines (Ly${\alpha}$, CIV$\lambda$1549, 
CIII]$\lambda$1909,  MgII$\lambda$2800, 
%[NeV]$\lambda\lambda$3346,3426,  
H${\beta}$ and H${\alpha}$) of width $>$2000~km/s are classified as broad-line AGNs. 
They include Type I Seyferts and quasars. These are sources, in which we
are able to have an unobscured view of their 
central nuclear region (Antonucci 1993). We detect 267 such 
objects (see Table 1) in our sample corresponding to 55\% of the identifications. 
A few examples are shown in Fig.~5. Redshifts are between 
0.15~$<$~$z$~$<$~3.87 with a median of $z_{\rm med}$~=~1.68 (see Fig.~2).

We also looked for the presence of Broad Absorption Line (BAL) quasars
in our spectroscopic sample. These BAL quasars are characterized by the 
presence in their spectra of strong absorption troughs blueshifted 
relative to the QSO emission redshift by 5000 to 50,000 km/s.
Seven of the XMM sources we have identified are BALs at $z >$ 1.5
based on the presence of strong CIV absorption. Further details on
these objects will be reported in a specific paper focussed only on BALs (Stalin 
et al. 2009a).

%\subsection{Narrow line AGN (NLAGN)}
%
%Objects with metal recombination and/or high ionization emission lines similar to those seen in BLAGN
%(such as CIV$\lambda$ 1549, CIII$\lambda$ 1909, MgII$\lambda$2800, [NeV]$\lambda\lambda$ 3346, 3426), 
%indicating the presence of an AGN, but with FWHM~$<$~2000 km/s are classified as Narrow line AGN.  
%There are xxx such sources in our sample. These are sources
%viewed edge on with an obscured view of the nucleus as per
%the unification scheme of AGN (Antonucci 1993).
%%For low redshift objects
%%(z < xxx) in which $H_{\alpha}$ was included in the AAT spectra
%%was used to  differentiate spectra that show narrow lines
%%due to ionization by hot stars from spectra that show narrow lines due
%%to an active nucleus.
%%Fig. xx shows the BPT diagram of objects with low z in our sample
%Lines such as Ly${\alpha}$, 
%%MgII $\lambda$ 2800, 
%H${\beta}$, H${\alpha}$, [OII]$\lambda$3727 etc. will also usually be present, but are alone not
%%sufficient to classify the source as an AGN.  
%%{\bf cite the high-z type II QSO of Mari et al. }

\subsection{Absorption line galaxies (ALGs)}
Objects visually identified by the 4000~\AA ~continuum break and absorption features such 
as  CaII-HK$\lambda\lambda$3934,3968 and with no obvious presence of  emission lines are 
classified as absorption-line galaxies. We do not impose
any limit on the equivalent width of emission lines.  
Some examples of observed ALGs are shown in Fig.~6.
We have identified 32 XMM sources as ALGs. They lie in the redshift range
$0.04 < z < 0.34$. The possible nature of this sample of ALGs is further
discussed in Section 5.2.

\begin{table}
% \centering
 \begin{minipage}{80mm}
\caption{Summary of the optical sources having two nearby XMM sources.
All optical sources  are found to be quasars}
\begin{tabular}{ccccc} \hline
 XMM$-$ID             & $\alpha_{2000}$  & $\delta_{2000}$  & Sep              & z  \\ 
                      &  (optical)       &    (optical)     & ($\prime\prime$) &   \\ \hline
J022046.1-042032 & 2:20:46.209 & -4:20:38.382 & 5.72 & 2.10  \\
J022046.2-042038 &             &              & 1.08 &        \\
J022245.5-041928 & 2:22:45.849 & -4:19:32.133 & 5.06 & 1.27  \\ 
J022246.0-041932 &             &              & 3.29 &        \\ 
J022336.8-034500 & 2:23:37.053 & -3:44:59.698 & 2.81 & 1.33  \\ 
J022337.3-034458 &             &              & 4.17 &        \\
J022339.0-042005 & 2:23:39.275 & -4:20:05.083 & 3.94 & 0.46 \\ 
J022339.3-042000 &             &              & 5.08 &        \\
J022624.5-041959 & 2:26:24.641 & -4:20:02.212 & 2.84 & 2.23 \\  
J022624.7-042005 &             &              & 3.93 &        \\       
J022643.5-041629 & 2:26:43.937 & -4:16:27.062 & 5.77 & 0.23 \\  
J022643.9-041625 &             &              & 1.21 &        \\  
J022647.1-041038 & 2:26:47.516 & -4:10:37.350 & 5.57 & 2.08 \\ 
J022647.5-041035 &             &              & 2.14 &        \\
J022733.9-042224 & 2:27:34.206 & -4:22:28.534 & 5.69 & 1.14 \\ 
J022734.2-042227 &             &              & 1.97 &        \\ \hline

\end{tabular}
\end{minipage}
\end{table}

\subsection{Emission line galaxies (ELGs)}
Sources with narrow emission lines, but with no obvious AGN features
in their optical spectra (e.g. high ionization and/or broad lines)
are classified as emission-line galaxies (ELGs). 
Examples of ELG spectra are shown in Fig.~7. Typical emission lines are
OII$\lambda$3727, H${\beta}$, [OIII]$\lambda$4959,5007, H${\alpha}$ etc.
Other features include CaII-H-K$\lambda\lambda$3934,3968 absorption, 
the continuum break at $\sim$4000~\AA, and narrow [NeIII]$\lambda$3869 emission. 
We have identified 116 ELGs in the redshift range 0.02 $<$ $z$ $<$ 0.76
(median $z_{\rm med}$~=~0.26).

%The emission lines in these 
%spectra indicate that the ionization mechanism inducing the optical spectrum
%is from hot stars not from a hard power law source.
This classification does not rule out the presence of some underlying AGN 
activity in ELGs. 
%Indeed, we believe that the majority of these objects present some level of AGN activity.   
The dominant energy source (AGN or starburst), can be 
identified using the commonly
used BPT diagnostic diagrams (Baldwin, Phillips, Terlevich 1982). 
The ratios of nebular emission lines ([OIII]$\lambda$ 5007/H${\beta}$) and 
([NII]$\lambda$ 6593/H${\alpha}$) are used to distinguish the ionising 
source i.e., between thermal continuum from starbursts and non-thermal 
AGN continuum. 
The BPT diagram for our sample of ELGs is shown in Fig.~8 together with
the two empirical relations commonly used to classify the 
emission-line objects
into AGNs (points above the curves) or starbursts (below the curves):

\begin{equation}
{\rm log} ([OIII]/H{\beta})  = \frac{0.61}{[{\rm log}([NII]/H{\alpha}) - 0.05]} + 1.3
\end{equation}

\begin{equation}
{\rm log}([OIII]/H{\beta})  = \frac{0.61}{[{\rm log}([NII]/H{\alpha}) - 0.47]} + 1.19
\end{equation}

Eq.~1 is given by Kauffmann et al. (2003) and Eq.~2 by Kewley et al. (2001).
%They are shown respectively as dotted and solid lines in Fig. ~7.
%In the following, we use the Kauffmann et al. (2003) relation.
%to separate the AGNs from the starburst galaxies. 
However, there are sources for which we are not able to 
measure both the [NII]/H${\alpha}$ or the [OIII]/H${\beta}$ line ratio.  
In such cases, an ELG is thought to be powered by an AGN if it has {\rm log}([NII]/H${\alpha}$)~$>$~$-$0.2
or {\rm log}([OIII]/H${\beta}$)~$>$~0.9 and by a starburst otherwise if we take
Kauffmann et al. (2003) relation. On the other hand, if we consider
Kewley et al. (2001) relation an ELG is thought to be powered 
by an AGN if it has {\rm log}([NII]/H${\alpha}$)~$>$~0.3 or {\rm log}([OIII]/H${\beta}$)~$>$~1.0. 
Based on the above
line ratio diagnostics and following Kauffmann et al. (2003), of the 116 emission-line galaxies, 
61 are powered significantly by
starbursts and the remaining 55 are significantly powered by 
AGN (so about 50\%). Alternately
if we consider Kewley et al. (2001) relation, in the sample of 116 ELGs, 82 are
powered by starbursts and 31 are powered by AGN. 
Another criteria 
to separate
the ELGs with X-ray emission dominated by stellar processes rather than
AGN activity is to  look into their X-ray to optical flux ratio ($f_x/f_o$; 
Fiore et al. 2003 (see Figure 5)).
If $f_x/f_o$ $<$ 0.01 it is a normal star-forming galaxy and if 
$f_x/f_o$ $>$ 0.1 it is an AGN (Kim et al. 2006). We define log ($f_x/f_o$) as
\begin{equation}
log (f_x/f_o) = log (f_x) + 5.41 + \frac{g^{\prime}}{2.5}
\end{equation}
Here $f_x$ is the flux in the 0.5$-$2keV band and $g^{\prime}$ is the optical
magnitude. The constant comes from the conversion of AB $g^{\prime}$ magnitude
into monochromatic flux and integration of the monochromatic flux over 
the $g^{\prime}$  bandwidth assuming a flat spectrum. Fig.~9  
shows $f_x/f_o$ versus the intrinsic X-ray luminosity in 
the 0.5$-$2 keV  band, log (L$_{(0.5-2keV)}$) for the broad-line AGNs, the two 
classes
of ELGs found from the BPT diagram (those significantly powered by AGN and
starburst) and ALGs. From the figure it is clear that the broad-line AGNs
in our spectroscopic sample occupy a distinct location in the
log  ($f_x/f_o$) v/s log (L$_{(0.5-2keV)}$) plane. ELGs and ALGs of our sample 
mostly occupy the region with intrinsic X-ray luminosity 
$L_{0.5-2keV} < 10^{43}$ erg s$^{-1}$. 
%Even if the consider the ELGs below the Kaufmann et al. (2003) curve in
%Fig.~7, it is difficult to identify what fraction of these belong to
%normal star-forming galaxies based on their location in 
%the log  ($f_x/f_o$) v/s log L$_{(0.5-2keV)}$ plot. 
%From Fig.~8 
It thus seems that 
at $L_{0.5-2keV} < 10^{43}$ erg s$^{-1}$ the ELGs population 
in our sample is made of 
%our sample of ELGs 
%could consist of a mixed population of sources such  
a mixture of normal star-forming galaxies, low-luminosity AGNs
 and galaxies powered by starbursts with high star-formation rates.
%whereas some ALGs could be giant elliptical galaxies with large amounts of 
%hot ISM (Kim et al. 2006).
In all further discussions, we 
however, use the Kauffmann et al. (2003) relation to separate ELGs 
%in AGN powered and starbursts powered systems. 
powered predominantly by AGN and starburst respectively.

\subsection{Stars}
As in other deep X-ray surveys, our sample also contains about 15\% (74 objects)
of galactic stars as optical counterparts to XMM sources (at $z$ = 0).
They are typically G, K and M-type stars whose
X-ray emission is caused by magnetic activity (Brandt \& Hasinger 2005). A few spectra
are shown in Fig.~10.

\subsection{Multiple optical counterparts}
In our spectroscopic sample, there are 8 cases for which two XMM sources have 
the same optical counterpart within 6 arcsecs of 
each of the XMM sources. The details of these 8 sources are given in Table 3. In
all these cases the optical counterpart is a quasar. 
%It is probable that the counterpart of the
%second source is much fainter.
We find also 16 XMM sources having more than one optical counterpart within
6 arcsecs of them. Spectra of all possible 
counterparts of the 16 XMM sources were obtained and all are classifed 
based on their spectral appearance. The details of these sources are shown 
in Table 4. Of these,
the two XMM sources, XLSS J022600.1$-$035955 and XLSS J022630.7$-$050550, 
are associated with a quasar-galaxy pair. It is more than probable 
that the XMM source counterpart is the quasar. 
%two optical sources each corresponding to the XMM sources, 
%XLSS J022600.1-035955, XLSS J022630.7-050550,  respectively are found to be 
%quasar galaxy pairs. 
Note that XLSS J022536.4$-$050011 is also associated with a radio source in 
the NVSS with a 1.4 GHz flux density of 12 mJy (see Table 5).

\subsection{Multiwavelength counterparts}
\subsubsection{Correlation with radio surveys}
We cross-correlated the 489 newly identified XMM sources with the publicly available
National Radio Astronomical Observatory VLA sky survey (NVSS; Condon
et al 1998) at 1.4 GHz. Only 21 XMM sources are
detected in NVSS. We also cross-correlated our 489 identified sources with
low frequency radio observations performed in this field with GMRT at 240 and
610 MHz (Tasse et al. 2007). There are 12 objects detected
at 240 MHz, while another 22 objects are detected
at 610 MHz. In total 34 objects are detected in one of the radio bands.
%Radio spectral indexes are determined
%for sources having flux measurements in two or more radio
%frequencies.
%, by linear fits to the data points in thelog Flux vs. log frequency plane.
The details of the XMM sources with radio detections are given in Table 5.
%{\bf PPJ: Give units in the Table. Give the radio indexes ???}.

\subsubsection{Correlations with IR surveys}
There is some overlap between the XMM-LSS field and one of the fields
observed by the {\it Spitzer} Wide-area Infrared Extragalactic Survey (SWIRE; Lonsdale
et al. 2003). In SWIRE, observations were performed with
the Infrared Array Camera (IRAC) at 3.6, 4.5, 5.8 and 80 $\mu$m and with
the Multiband Imaging Photometer (MIPS) at 24, 70 and 160 $\mu$m to a 
5$\sigma$ depth of 4.3, 8.3, 58.4, 65.7 $\mu$Jy and 0.24, 15 and
90 mJy respectively (Tajer et al. 2007). Correlating our sample of 
optically identified XMM sources (489 in total) with SWIRE, we find that 
$\sim$50\% (239 sources) are detected in SWIRE.  Of these, 133 are
broad-line AGNs, 36 are stars, 50 and 20 are emission and absorption-line
galaxies respectively. The positions of these X-ray sources 
in the IR colour$-$colour diagram (3.6$-$4.5 v/s 5.8$-$8.0 $\mu m$) 
is shown in Fig.~11. Optical identification and description of 
SWIRE sources over
a larger area in CFHTLS, will be reported 
in Stalin et al. (2009b).

\section{Discussion}
%
%Among the spectroscopically classified sources, we find 225 
%BLAGN ($\sim \%$ of the sample), xxx NLAGN ($\sim \%$ of the sample), 
%113 ELG ($\sim \%$ of the sample), 34 ALGs ($\sim \%$ of the sample), and 71 
%stars ($\sim 15\%$ of the sample). Among the 113 ELGs and based on BPT diagnostic 
%diagram, 56 are found to be AGNs and 57 to be powered by starbursts. 
%A summary of our identifications is given in Table 1. Table 2 gives the list
%of XMM sources which are identified as BLAGN. The list of indentified
%narrow line objects (including NLAGN and ELGs) are given in Table 3. List of 
%absorption-line galaxies and stars are given in Tables 4 and 5 respectively.

%\begin{figure}
%\psfig{file=Na5.ps,width=9cm,height=9cm}
%\caption{Plot of $\alpha_{\rm ox}$ as a function of the rest frame optical 
%luminosity at 2500 \AA. Symbols are the same as in Fig. 4 and the solid line is the
%linear least squares fit to the broad line objects.}
%\end{figure}

%\begin{figure}
%\psfig{file=Na6.ps,width=10cm,height=10cm}
%\caption{Plot of $\alpha_{\rm ox}$ vs the rest frame monochromatic
%luminosity at 2 keV. Symbols have the same meaning as in Fig. 4. Linear
%least squares fit to the broad line objects is shown as a solid line.}
%\end{figure}

\subsection{Optical-to-X-ray slope ($\alpha_{\rm ox}$)}

The broad band spectral index $\alpha_{\rm ox}$ of any source characterizes the UV to X-ray 
spectral energy distribution by assuming that the rest frame flux emitted at 2500~\AA~
can be connected to the one at 2~keV with a simple power law. This is a simple measurement 
of the amount of X-ray radiation emitted mostly by non-thermal processes with respect to the
amount of UV radiation emitted mostly by thermal processes (Kelly et al. 2007).
We estimated $\alpha_{\rm ox}$ for each of the spectroscopically identified 
broad-line AGNs and ELGs (as some of the ELGs are powered by AGN) in our sample. For this 
we have converted the observed $i^{\prime}$-band magnitudes to fluxes following the definition
of the AB system (Oke \& Gunn 1983)
%
%{\bf PPJ: Please check all these equations and define better the letters. Example:
%in Eq4, the ratio nu1/nu2 should be at the power $\alpha_0$ not at the power 2. Also
%S should be defined and nu1 as well because I think F is the rest frame luminosity
%so that nu1=2500A no ???? Also, S is used in Eq4 when F in equation 5 when the text
%says that's the same equations ?????? }

\begin{equation}
S_{\rm i^{\prime}} = 10^{-0.4 (m_{\rm i^{\prime}} + 48.60)}.
\end{equation}
where $S_{\rm i^{\prime}}$ and $m_{\rm i^{\prime}}$ are respectively the flux and 
magnitude in the i$^{\prime}-$band.
The luminosity at the frequency corresponding to 2500~\AA~ in the rest-frame is calculated 
following Stern et al. (2000)
\begin{equation}
L_{\nu_1} = \frac{4 \pi D_{\rm l}^2}{(1+z)^{1+\alpha_{\rm o}}} (\frac{\nu_1}{\nu_2})^{\alpha_o} S_{\nu_2}
\end{equation}
where $\nu_2$ is the observed frequency corresponding to 
the $i^{\prime}-$band,  $S_{\nu_2}$ is the observed flux 
in $i^{\prime}-$ band, $\nu_1$ is the rest-frame frequency corresponding 
to 2500 \AA ~and $D_l$ is the luminosity distance. An optical spectral 
index $\alpha_{\rm o}$ = $-$0.5 (Anderson et al. 2007) is 
assumed ($S_{\nu}$~$\propto$~$\nu^{\alpha}$). 
The luminosity at rest frame 2~keV is obtained using a similar equation as Eq.~5,
assuming  a X-ray spectral index of $\alpha_{\rm x}$ = $-$1.5 (Anderson et al. 2007).  

%
%\begin{equation}
%L_{\rm X} = \frac{4 \pi D_l^2}{(1 + z)^{1+\alpha}} F_X
%\end{equation}
%%
%
Thus, the broad band spectral index $\alpha_{\rm ox}$ is obtained as
\begin{equation}
\alpha_{\rm ox} = \frac{{\rm log}(L_X/L_{opt})}{{\rm log}(\nu_X/\nu_{opt})}
\end{equation}
Here $L_X$ and $L_{opt}$ are the rest frame monochromatic luminosities
at 2 keV and 2500\AA ~(erg s$^{-1}$ Hz$^{-1}$) respectively. The distribution of $\alpha_{\rm ox}$ 
(not corrected for
intrinsic and galactic absorption) for the sample of spectroscopically
identified broad-line AGNs is shown in Fig.~12. 
%for the different classes 
%of objects. 
%Emission and absorption-line galaxies are found to have a mean value of 
%$\alpha_{\rm ox}$ $-$1.90$\pm$0.10 and $-$2.23$\pm$0.08 respectively, 
%whereas stars are found to have a mean $\alpha_{\rm ox}$ of $-$2.46$\pm$0.09. 
A mean value of $\alpha_{\rm ox}$~=~$-$1.47$\pm$0.03 is found 
for broad-line AGNs 
which is consistent with the values found in the literature
(Strateva et al. 2005). 
%Starburst and AGN powered emission-line galaxies have a mean value of 
%$\alpha_{\rm ox}$ of $-$1.86$\pm$0.09 and $-$1.96$\pm$0.10 respectively.
%{\bf PPJ: You could also comment on the AGN-type galaxies compared
%to the Star-burst like galaxies (separation using the BPT diagrammes may be in
%an additional paragraph.}.

\begin{figure}
\hspace*{-0.5cm}\psfig{file=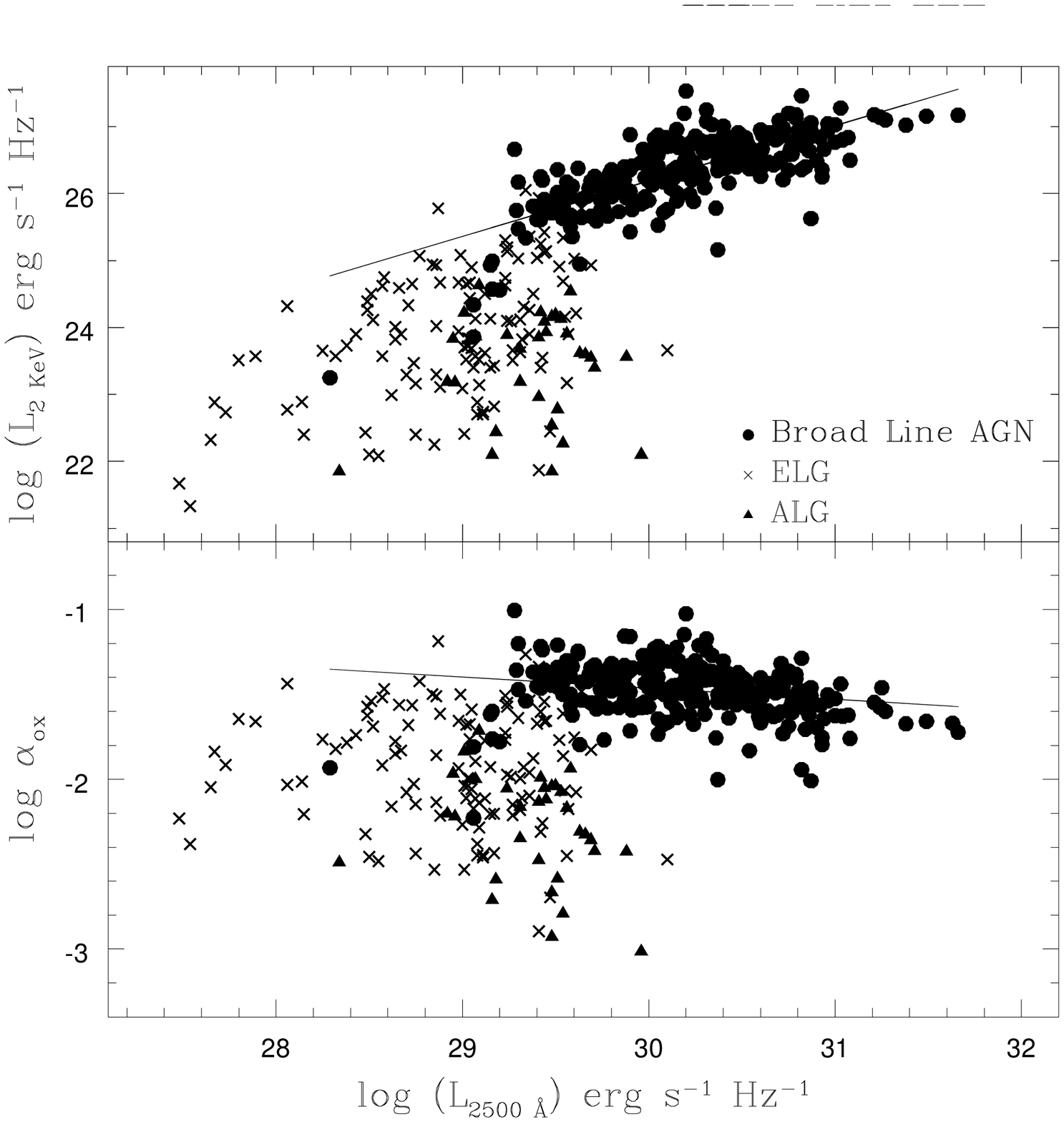,width=9cm,height=9cm}
\caption{{\it Top panel}: Plot of the rest frame monochromatic luminosity 
at 2 keV versus the rest frame monochromatic luminosity at 2500~\AA. 
{\it Bottom panel}: plot of $\alpha_{ox}$ against the rest frame 
monochromatic luminosity at 2500~\AA. Symbols have the same meaning 
as in Fig.~4. The solid line is the linear least squares fit to only
the broad-line AGNs}
\end{figure}

\begin{figure}
\hspace*{-0.6cm}\psfig{file=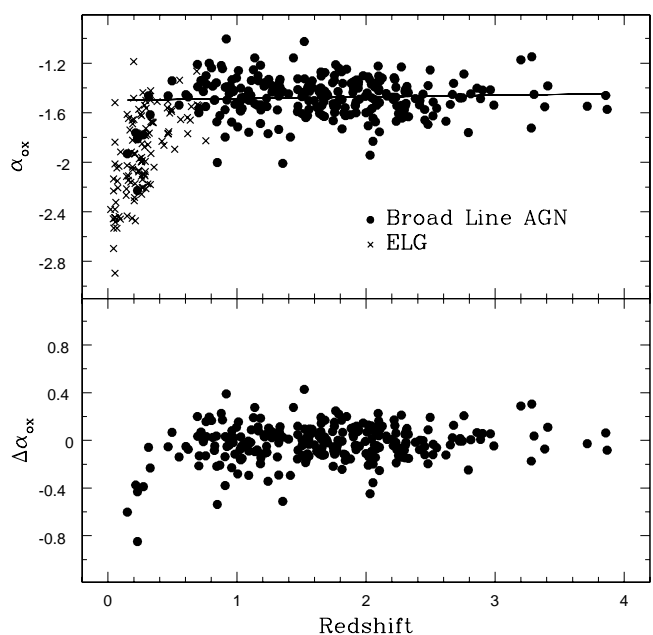,width=9cm,height=9cm}
\caption{{\it Top panel}: $\alpha_{\rm ox}$ versus redshift for 
broad-line AGNs and ELGs (as some of the ELGs are powered by AGN).
The solid line is the linear least squares fit to only the broad-line AGNs.
{\it Bottom panel}: Residual in $\alpha_{ox}$ plotted against 
redshift for only broad line AGNs. Symbols are the same as in Fig. 4.} 
\end{figure}

\begin{table*}
\centering
 \begin{minipage}{140mm}
\caption{Summary of XMM sources having more than one optical counterpart}
\begin{tabular}{cccccccc} \hline
XMM$-$ ID             & $\alpha_{2000}$ & $\delta_{2000}$ & Sep. & $g^{\prime}$         & ID  &  $z$ \\ 
                      & (optical)       & (optical)       & ($\prime\prime$) & (mag) &  \\ \hline
XLSS J022049.4-043030 & 02:20:49.545 & -04:30:28.774 & 1.77 & 21.99 & QSO  & 1.04 \\ 
                      & 02:20:49.501 & -04:30:31.265 & 1.27 & 21.47 & QSO  & 1.81 \\ 
XLSS J022105.4-044101 & 02:21:05.601 & -04:41:01.490 & 1.80 & 18.31 & ELG  & 0.20 \\ 
                      & 02:21:05.508 & -04:41:03.581 & 1.63 & 20.16 & ELG  & 0.20 \\ 
XLSS J022124.2-042517 & 02:21:24.264 & -04:25:20.231 & 2.75 & 19.28 & STAR & 0.00 \\ 
                      & 02:21:24.492 & -04:25:17.663 & 3.83 & 20.25 & ELG  & 0.29 \\ 
XLSS J022127.7-043407 & 02:21:27.978 & -04:34:10.234 & 4.38 & 17.20 & ALG  & 0.21 \\ 
                      & 02:21:27.567 & -04:34:03.263 & 5.01 & 20.06 & ELG  & 0.09 \\ 
XLSS J022159.6-045738 & 02:21:59.985 & -04:57:36.175 & 5.61 & 18.45 & ELG  & 0.16 \\ 
                      & 02:21:59.581 & -04:57:38.803 & 1.08 & 21.46 & ELG  & 0.16 \\ 
XLSS J022249.7-044539 & 02:22:49.768 & -04:45:36.907 & 2.11 & 19.95 & ELG  & 0.26 \\ 
                      & 02:22:49.835 & -04:45:39.856 & 1.38 & 20.52 & ELG  & 0.26 \\ 
XLSS J022316.0-040503 & 02:23:16.233 & -04:05:08.573 & 5.75 & 18.74 & STAR & 0.00 \\ 
                      & 02:23:16.091 & -04:05:03.988 & 0.97 & 21.56 & QSO  & 2.02 \\ 
XLSS J022329.1-045453 & 02:23:28.996 & -04:54:54.474 & 3.16 & 21.40 & ELG  & 0.33 \\ 
                      & 02:23:29.253 & -04:54:51.875 & 1.59 & 20.43 & QSO  & 0.60 \\ 
XLSS J022416.8-050325 & 02:24:16.997 & -05:03:24.466 & 2.56 & 18.27 & ALG  & 0.14 \\ 
                      & 02:24:16.863 & -05:03:24.993 & 0.64 & 19.21 & ALG  & 0.14 \\ 
XLSS J022424.8-052037 & 02:24:24.896 & -05:20:36.892 & 0.73 & 19.64 & ELG  & 0.28 \\ 
                      & 02:24:24.984 & -05:20:40.063 & 3.25 & 21.71 & ELG  & 0.28 \\ 
XLSS J022536.4-050011 & 02:25:36.438 & -05:00:11.922 & 1.00 & 16.00 & ELG  & 0.05 \\  
                      & 02:25:36.774 & -05:00:07.898 & 5.43 & 17.90 & ELG  & 0.05 \\ 
                      & 02:25:36.289 & -05:00:07.697 & 4.80 & 18.96 & ELG  & 0.05 \\ 
XLSS J022558.7-051245 & 02:25:59.150 & -05:12:45.378 & 5.92 & 19.35 & ELG  & 0.26 \\ 
                      & 02:25:59.036 & -05:12:47.268 & 4.49 & 21.94 & ELG  & 0.26 \\
XLSS J022600.1-035955 & 02:25:59.881 & -03:59:59.003 & 5.06 & 18.28 & ELG  & 0.14 \\
                      & 02:26:00.266 & -03:59:52.683 & 3.60 & 17.32 & STAR & 0.00 \\
                      & 02:26:00.017 & -03:59:54.375 & 2.29 & 21.07 & QSO  & 1.09 \\
XLSS J022630.7-050550 & 02:26:30.742 & -05:05:44.427 & 5.70 & 20.26 & ELG  & 0.19 \\
                      & 02:26:30.777 & -05:05:52.004 & 1.90 & 19.86 & QSO  & 1.66 \\  
XLSS J022635.1-050518 & 02:26:35.031 & -05:05:20.535 & 3.09 & 16.87 & STAR & 0.00 \\ 
                      & 02:26:35.259 & -05:05:17.719 & 1.58 & 17.03 & STAR & 0.00 \\ 
XLSS J022647.7-041426 & 02:26:47.693 & -04:14:30.763 & 4.44 & 20.78 & QSO  & 2.33 \\ 
                      & 02:26:47.848 & -04:14:26.229 & 1.48 & 21.05 & ELG  & 0.26 \\  \hline

\end{tabular}
\end{minipage}
\end{table*}

\begin{table*}
%\centering
%\hspace*{-1.0cm}{
\begin{minipage}{180mm}
\caption{Summary of XMM sources with radio detections }
\begin{tabular}{cccclllcrrr} \hline
XMM$-$ID              & $\alpha_{2000}$ & $\delta_{2000}$ & Sep & ~~~$g^{\prime}$  & ~~~z     & ID  &  S$_{240 MHz}$  & S$_{610 MHz}$  & S$_{1420 MHz}$  \\     
                      &  optical       &  optical     & (arcsec)& (mag)   &          &     & (mJy)~~~   & (mJy)~~~  & (mJy)~~~   \\ \hline
XLSS J022001.6-052217 &  02:20:01.624  & -05:22:16.904 &  0.70 & 20.12 &   2.22 & QSO &  9.6  &   15.4&   15.4 \\
XLSS J022108.2-042759 &  02:21:08.366  & -04:28:01.924 &  3.05 & 20.18 &   0.27 & ELG &  ---  &   --- &   2.1 \\
XLSS J022120.1-040217 &  02:21:20.477  & -04:02:17.911 &  4.54 & 20.31 &   0.32 & ALG &  ---  &   --- &   3.3 \\
XLSS J022127.7-043407 &  02:21:27.978  & -04:34:10.234 &  4.38 & 17.20 &   0.21 & ALG &  17.5 &   11.1&   8.3 \\
XLSS J022127.7-043407 &  02:21:27.567  & -04:34:03.263 &  5.01 & 20.06 &   0.09 & ELG &  17.5 &   11.1&   8.3 \\
XLSS J022144.9-035745 &  02:21:44.947  & -03:57:45.388 &  0.61 & 21.45 &   2.50 & QSO &  51.6 &  20.4 &   9.6 \\
XLSS J022247.7-043330 &  02:22:47.874  & -04:33:30.031 &  1.42 & 20.32 &   1.63 & QSO &  ---  &   2.4 &   4.5 \\
XLSS J022251.6-050714 &  02:22:51.748  & -05:07:12.422 &  2.40 & 21.53 &   3.86 & QSO &  ---  &   --- &   3.5 \\
XLSS J022255.5-051817 &  02:22:55.739  & -05:18:17.404 &  3.59 & 20.85 &   1.76 & QSO &  670.2&  436.9&  259.5 \\
XLSS J022257.9-041840 &  02:22:57.965  & -04:18:40.643 &  0.37 & 19.05 &   0.24 & ELG &  ---  &    2.8&  --- \\
XLSS J022258.4-040709 &  02:22:58.568  & -04:07:15.079 &  5.62 & 19.96 &   0.29 & ALG &   7.0 &    ---&  --- \\
XLSS J022310.2-042304 &  02:23:10.029  & -04:23:04.034 &  3.63 & 21.03 &   0.24 & ELG &  39.3 & 52.0  &  37.2 \\
XLSS J022337.4-040937 &  02:23:37.439  & -04:09:38.303 &  0.60 & 14.78 &   0.00 & STA &  ---  &  6.6  &   4.0 \\
XLSS J022345.3-043406 &  02:23:45.337  & -04:34:07.965 &  1.43 & 14.94 &   0.00 & STA &  ---  &  2.9  &   3.5 \\
XLSS J022402.4-044135 &  02:24:02.640  & -04:41:34.703 &  2.25 & 14.54 &   0.04 & ELG &  ---  &  ---  &   4.7 \\
XLSS J022403.7-043303 &  02:24:03.782  & -04:33:04.893 &  1.49 & 15.55 &   0.04 & ELG &  11.4 &   6.8 &    4.0 \\
XLSS J022447.0-040849 &  02:24:47.000  & -04:08:51.096 &  1.54 & 17.77 &   0.10 & ELG &  ---  &   1.8 &    --- \\
XLSS J022509.7-050950 &  02:25:09.712  & -05:09:49.145 &  1.48 & 20.20 &   0.32 & ALG &  ---  &   2.4 &    --- \\
XLSS J022528.3-041536 &  02:25:28.348  & -04:15:39.796 &  3.68 & 21.42 &   0.56 & ELG &  ---  &  ---  &    2.4 \\
XLSS J022536.4-050011 &  02:25:36.438  & -05:00:11.922 &  1.00 & 16.00 &   0.05 & ELG &  33.2 &   16.3&   12.0  \\
XLSS J022536.4-050011 &  02:25:36.774  & -05:00:07.898 &  5.43 & 17.90 &   0.05 & ELG &  33.2 &   16.3&   12.0 \\
XLSS J022536.4-050011 &  02:25:36.289  & -05:00:07.697 &  4.80 & 18.96 &   0.05 & ELG &  33.2 &   16.3&   12.0 \\
XLSS J022549.4-040028 &  02:25:49.766  & -04:00:24.460 &  5.79 & 16.27 &   0.04 & ELG &  ---  &    2.9&    --- \\
XLSS J022556.1-044724 &  02:25:56.092  & -04:47:24.403 &  0.84 & 21.43 &   1.01 & QSO &  ---  &    2.3&    2.4 \\
XLSS J022558.7-050055 &  02:25:58.853  & -05:00:54.245 &  2.24 & 18.58 &   0.15 & ELG &  7.8  &    6.9&    3.2 \\
XLSS J022604.3-045929 &  02:26:04.510  & -04:59:33.441 &  4.83 & 16.21 &   0.05 & ELG &  ---  &    3.3&    --- \\
XLSS J022609.9-045805 &  02:26:09.666  & -04:58:05.615 &  3.68 & 15.35 &   0.05 & ALG &  ---  &    2.1&    --- \\
XLSS J022617.6-050443 &  02:26:17.408  & -05:04:43.312 &  3.47 & 15.50 &   0.05 & ELG &  20.6 &   13.1&    8.4 \\
XLSS J022618.9-040016 &  02:26:19.060  & -04:00:14.727 &  2.47 & 19.85 &   0.21 & ALG &  118.8&   87.8&    --- \\
XLSS J022659.2-043529 &  02:26:58.959  & -04:35:26.615 &  5.29 & 14.74 &   0.07 & ALG &  ---  &   --- &    3.2 \\
XLSS J022720.7-044537 &  02:27:20.692  & -04:45:37.166 &  0.89 & 15.98 &   0.05 & ELG &  6.7  &    3.2&    --- \\
XLSS J022740.5-040250 &  02:27:40.545  & -04:02:50.990 &  0.72 & 18.18 &   2.62 & QSO &  ---  &   3.3 &    3.3 \\ 
XLSS J022758.1-040753 &  02:27:57.999  & -04:07:51.950 &  2.16 & 19.43 &   0.21 & ELG &  ---  &   --- &    2.6 \\ \hline
\end{tabular}
\end{minipage}
%}
\end{table*}

\subsubsection{X-ray luminosity (L$_{X}$) versus optical luminosity (L$_{opt}$)}

Previous studies of optically-selected quasars have found a strong correlation between the 
optical and X-ray monochromatic luminosities (Anderson et al. 2007 and references therein).
The relation between both luminosities is of the form $L_{\rm X} \propto L_{\rm opt}^{\beta}$.
Different values have been derived for $\beta$ varying typically from 0.7 (Pickering et al. 1994; 
Wilkes et al. 1994) to 0.8 (Avni \& Tananbaum 1986) and up to unity or slighly larger 
(La Franca et al. 1995; Green et al. 2009). The plot giving the rest frame monochromatic 
luminosity at 2~keV versus the rest frame optical luminosity at 2500~\AA~ of all the optically 
identified X-ray sources is shown in Fig.~13 (top panel). It is apparent that a strong
correlation exists for broad-line AGNs. Part of the scatter in the X-ray to optical relation 
in Fig.~13 (top panel) might be related to the variability of the AGNs as the X-ray and optical observations are
far from being simultaneous, but we do not expect this effect to be 
very important. The fit for only the broad-line AGNs gives

\begin{equation}
{\rm log} L_{\rm X} = (0.870 \pm 0.001){\rm log} L_{opt} + 0.009 \pm 0.370 
\end{equation}

%AGNs with intrinsic absorption that 
%can affect UV {\bf PPJ: You said you did not correct for absorption... how did
%you determined these objects ???? That cannot be only BALs}
%and X-ray photons differently and 
with a linear correlation coefficient $r$~=~0.98. This is 
slightly lower than the slope of 1.12 found by Green et al. (2009), however, within the range of values found by Steffen et al. (2006) through
different regression methods.  Note that to perform the fit we excluded 
radio loud AGNs that can have additional 
UV and X-ray flux associated with the radio jet (Worrall et al.  1987; Worrall \& Birkinshaw 2006;
Wilkes \& Elvis 1987). Also, radio-loud AGNs are found to be 2$-$3 times
brighter in X-ray for the same optical magnitude (Shen et al. 2006; 
Zamorani et al. 1981).  Therefore we have used only the sample of broad-line 
AGNs, excluding the 9 AGNs which are found to be 
detected in the radio (see Table. 5).

\subsubsection {$\alpha_{\rm ox}$ versus optical luminosity ($L_{opt}$)}
The existence of a correlation between $L_{X}$ and $L_{opt}$ 
in broad-line AGNs, in turn implies that $\alpha_{\rm ox}$
correlates also with the X-ray and UV monochromatic luminosities at 2 keV and
2500~\AA~ respectively. Fig.~13(bottom panel) shows the trend 
of $\alpha_{\rm ox}$ with 
the rest-frame monochromatic luminosity for our spectroscopic sample.
For broad-line AGNs, from a linear least squares fit we found 
\begin{equation}
\alpha_{\rm ox} = (-0.065 \pm 0.019) log (L_{opt}) + (0.509 \pm 0.560)
\end{equation}
This is similar to the value of 0.060 $\pm$ 0.007 found by Green et al. (2009)
however, flatter than the value of 0.137 $\pm$ 0.008 found by Steffen
et al. (2006). We do not find any significant correlation between 
$\alpha_{\rm ox}$ and ($L_{opt}$) for our sample of ELGs.

\subsubsection{Evolution of $\alpha_{\rm ox}$}
%The broad-line AGN in our spectroscopic sample show correlation between
%$\alpha_{\rm ox}$ ($L_{2500 \AA}$) , $\alpha_{\rm ox}$ ($L_{0.5-2keV}$) 
%and also 
%show a luminosity redshift trend. Because of the above correlations, $\alpha_{\rm ox}$ must show some correlation with redshift as well.
The plot of $\alpha_{\rm ox}$ versus 
%of the identified X-ray sources 
redshift is shown in Fig.~14(top panel). 
The fit to the data in our sample of broad-line AGNs
with $\alpha_{\rm ox} = A z + B$
gives A~=$-$0.021$\pm$0.015 and a linear correlation coefficient of 0.07. 
It thus seems  that for broad-line AGNs, $\alpha_{\rm ox}$ 
does not depend strongly on redshift. 
This is consistent with the claims of non-evolution already
published in the literature (Just et al. 2007; Steffen et al. 2006; Vignali et al. 2003; 
Strateva et al. 2005; Avni \& Tananbaum 1986;  Green et al. 2009; however
see Yuan et al. 1998; Bechtold et al. 2003).
We also tried to check for the evolution of $\alpha_{\rm ox}$ for broad-line AGNs after removal of its luminosity dependence. For this we subtracted
the $\alpha_{\rm ox}$ obtained through the best fit $\alpha_{\rm ox}$$-$
$L_{opt}$ regression (Eq.~8) from the observed  $\alpha_{\rm ox}$. We 
find that this resultant $\Delta \alpha_{\rm ox}$ for broad line AGNs shows no trend with 
redshift (Fig.~14 bottom panel). 
%From Fig.~14(top panel)  it is seen that at 
%$z<0.5$, $\alpha_{\rm ox}$ shows a trend with {\it z}.  
%We have only 7 broad-line AGN at {\it z} $<$ 0.5 (this is not sufficient to 
%look for any such correlation among broad-line AGN)  and
%majority of the other sources at {\it z} $<$ 0.5 are ELGs and from BPT analysis
%about 50\% of them are powered significantly by AGN. 
%This evolution of $\alpha_{ox}$ with {\it z} at {\it z} $<$ 0.5, might 
%not be real and has to be taken with caution, as these low 
%redshift ELG sources are characterized in 
%their optical spectrum by the presence of a significant/dominat 
%contamination of star-light from the host galaxy. 
%This may be due to the fact that we pick up
%objects intrinsically much weaker than at higher redshift (see also Yuan et al. 1998 and 
%Bechtold et al. 2003). Majority of the objects at {\it z} $<$ 0.5 are ELGs and from
%BPT diagram $\sim$50\% of them are powered by AGN.
% {\bf PPJ: Please check}.
%
%The only results indicating a dependence of $\alpha_{\rm ox}$ on
%{\it z} is by Yuan et al. (1998), however at $z$ $<$ 0.5  and Bechtold et al. (2003). 
%This might be due to the heterogeneous nature of their sample.
%{\bf may be you should see whether Yuan et al's and Bechtold et al.'s correlation is
%due to small z range couvered close to z=0. In the figure if you
%restrict to z<1.5 you may get a weak trend.}
It can be be seen in Fig. 14 that for {\it z} $<$ 0.5, (i) most of the 
$\alpha_{\rm ox}$ measurements are smaller than $-$1.6 and (ii) the scatter
in the values of $\alpha_{\rm ox}$ is much larger than for higher redshifts.
The main reason for this is that the low redshift sources are mostly ELGs (50\%
of them being significantly powered by AGN). The optical spectrum of
these objects firstly is significantly/predominantly contaminated by star-light
from the host galaxy and secondly, the derivation of $\alpha_{\rm ox}$
using the optical and X-ray spectral index of AGNs might not be appropriate
in ELGs.

An ideal sample to look for the correlations between $L_X$ and
$L_{opt}$, and  $\alpha_{ox}$ with either $L_X$, $L_{opt}$ or {\it z} is
the one that fills the $L-z$ plane. However, in practice it is
difficult and most of the samples in literature used for such correlation
searches were based on objects that covers only a narrow range in the
$L-z$ plane. Such a sample is also likely to be contaminated with
sources with unrelated physical processes such as BALs and radio-loud AGN
(Green et al. 2009).  %The usual practice is also to eliminate from the sample 
%BALs and radio-loud quasars to minimize
%contamination of pure accretion dominated X-ray emission. 
We wish to point out that the linear least squares regression analysis
presented here is a simple minded approach compared to the detailed 
statistical
analysis performed by Green et al. (2009) that includes upper limits.
However as discussed in earlier sections, our results are broadly 
in line with those available in literature.

%------------------
%However, our results for $\alpha_{OX}$ v/s optical luminosity is consistent with Green 
%et al. (2009). Moreover our resuls for the linear regression coefficient between $L_{X}$ 
%and $L_{opt}$ is also consistent with Green et al. (2009) if we restrict only to the 
%sample of Green et al. (2009) whcih exclues BALs RAdio loud quasars. }
%---------------------

%could be attributed to the effect of upper 
%limits (such as X-ray detected AGN without identifications,
%or optical magnitudes falling below the flux limit of our this spectroscopic
%survey) that are not considered in our analysis.}

%We caution on the impact, the sample selection could have
%in the various correlations reported here. These correlations,  were based on
%simple linear least squares regression analysis and do not take into account
%objects with upper limits, such as X-ray detected AGN without identifications,
%or optical magnitudes falling below the flux limit of our this spectroscopic
%survey. Such selection biases can be alleviated by selection of appropriate
%sub-samples as well as carrying our different statistical tests like
%the survival analysis as has recently been done by Green et al. (2009), which
%we do not try to do here. Similarly, in  studies
%available in literature, where attempts have been made to find these
%correlations, BALs and radio-loud quasars will be eliminated to minimize
%contamination of pure accretion dominated X-ray emission. We too have adopted
%this in our work, however, this might also induce some biases, which has
%been nicely pointed out recently in Green et al. (2009).}

\subsection{The absorption-line galaxies}

The existence of an intriguing population of galaxies with strong X-ray
emission
(41 $<$ log L$_X$ $<$ 44 erg s$^{-1}$  in the rest frame energy range
2$-$8 keV; Rigby et al. 2006)
was first noted from observations with the Einstein Observatory (Elvis
et al. 1981).
The X-ray emission reveals the presence of an AGN without other sign of
activity
(no emission line is seen in their optical spectrum). Further
observations with
ROSAT, {\it Chandra} and {\it XMM-Newton} have supported the existence
of this population
(Fiore et al. 2003; Caccianiga et al. 2007; Griffiths et al. 1995;
Garcet et al. 2007).
Such galaxies are now referred to as X-ray Bright Optically Normal Galaxies
(XBONGs). 

Their exact nature is unknown although several
hypothesis have been advocated in the literature to explain their
properties.
It has been argued that heavy obscuration by Compton-thick gas covering
the nuclear source
could prevent ionizing photons from escaping. In this hypothesis, the
obscuration must occur in all directions, and should not be restricted
to the torus, because
the sources lack both broad and narrow lines in their spectra. This idea
might be correct for at least some XBONGs (Spoon et al. 2001).
Dilution of nuclear emission from the host galaxy starlight could play a
role(Georgantopoulos
\& Gerogakakis 2005). Such an effect might be important if the ground based
spectroscopic observations are performed with relatively wide slits.
Severgnini et al. (2003) have shown on three objects of their sample, 
that adequate observations (narrow slit, accurate positioning of the
slit etc.) reveal
the missing emission lines. They argue that
survey observations are sometimes inadequate to completely unveil the true
nature of the X-ray sources. However, the intrinsic optical continuum
emission
of XBONGs is weak compared to that of quasars (Comastri et al. 2002)
suggesting that the emission line luminosities are also intrinsically
small.
XBONGs could be extreme BL Lacs in which the featureless non-thermal
continuum is much weaker than the host galaxy starlight. Indeed, there
is one case of
a XBONG being associated with a BL Lac object (Brusa et al. 2003) and it
has been argued by Fosatti
et al. (1998) that XBONGs could belong to the low-luminosity tail of the
blazar sequence.
However, large calcium break and radio quietness
in several XBONGs argue against them being BL Lacs (Fiore et al. 2000).
Finally, XBONGs might be powered by an inner Radiatively
Inefficient Accretion Flow (RIAF) plus an outer radiatively efficient
thin accrection disk (Yuan \& Narayan 2004).

In our sample we detect 32 X-ray sources having only absorption lines
and devoid of any emission lines. This represent $\sim$6\% of
our sample of observed sources. They have
redshifts ranging between 0.04 and 0.34. These galaxies are located in the
peculiar regions of the IR-color-color diagram (see Fig.~11) well outside the AGN
location and have also a distinct location in the L$_{opt}$ $-$ L$_{X}$
plane (see Fig.~13). We further analysed the optical spectra of these
sources
to see if they could be BL Lacs by looking at the shape of the continuum
around the  Ca H\&K break at 4000 \AA. The detection of a significant
reduction
of the Ca H\&K break when compared to normal elliptical galaxies can be
considered as an indication of the presence of a substantial non-thermal
nuclear continuum emission. To identify the shape of the continuum around
4000 \AA ~we define D$_{\rm n}$ (relative flux depression across the Ca
H\&K break) as
\begin {equation}
D_{\rm n} = \frac{F^+ - F^-}{F^+}
\end{equation}

Here F$^+$ and F$^-$ are the average fluxes in the region 4050$-$4250\AA
~and
3750$-$3950\AA ~respectively in the source rest frame (Caccianiga et al.
2007).
We adopt a limit of $D_{\rm n}$ $<$ 40\% for a source with no emission line
to be a BL Lac object (Caccianiga et al. 2007). Of the 32 sources in our
sample
17 have $D_{\rm n}$ $<$ 40\% and hence could be BL Lac candidates. To
further investigate the possibility for these 17 sources to be BL Lac
candidates
we looked for any sign of radio emission from these objects (this is
another indication of the
presence of non-thermal emission) both in the literature and the NVSS.
Three among the 17 sources with $D_{\rm n}$ $<$ 40\% are found to be
radio emitters.
It is thus likely that these 3 objects (XLSS J022509.7$-$050950, XLSS
J022609.9$-$045805, XLSS J022659.2$-$043529)
are BL Lacs.
%The ratio $L_{\rm UV}/L_{\rm X}$ is also high compared to the ratios
%observed for
%the X-ray sources without broad emission lines (see Fig.~12).
It is also probable that emission lines in ALGs are weak and could be
revealed by deeper observations.
%but it seems that these objects represent a special population.
It would be interesting to conduct follow-up observations of our sample
of 32
absorption-line galaxies to study this population in more detail.
 
\section{Conclusions}
We have presented the optical spectroscopic classification of 489
X-ray sources. They were drawn from an initial sample of 829 X-ray sources
detected in the course of the XMM-LSS survey that overlap one of the wide
fields of the CFHTLS. The sources have a detection threshold
of 15 in either the 0.5$-$2 keV or 2$-$10 keV bands and are brighter
than 22 mag
in the optical $g^{\prime}$-band. This sample of 829 X-ray sources is thus
flux limited in both X-ray and optical bands. We observed 695 of these
objects with
the AAOmega system at the AAT of which spectroscopic identification was
possible for
489 sources (our sample is therefore $\sim$70\% complete).
%We hope to observe the remaining targets in future spectroscopic
%observations.
A large fraction of these X-ray sources were identified
as BLAGNs ($\sim$55\%) based on the FWHM of the emission lines present
in their spectra. In addition to these BLAGNs, 74 sources are identified
as galactic stars, 32 are absorption-line galaxies, and 116 are
emission-line galaxies. Based on BPT diagram we find that the emission
lines in 55 (61) of the 116 emission line galaxies are powered
significantly by an
AGN (starburst). For the broad-line AGNs, $\alpha_{\rm ox}$ is
found to correlate
well with the rest frame monochromatic luminosity at 2500~\AA~ (see also
Krumpe et al. 2007,
Just et al. 2007, Gibson, Brandt \& Schneider 2008, Green et al. 2009).
The rest frame X-ray and UV monochromatic luminosities at 2 keV and
2500~\AA~ respectively
are found to be closely correlated for broad line AGNs.
%Also $\alpha_{\rm ox}$ for the sample of AGN is found to show a clear
%correlation with the rest-frame monochromatic
%luminosity at 2 keV.
No dependence of $\alpha_{\rm ox}$ with redshift is found.
This is similar to the results by Just et al. (2007) and Green et al.
(2009) but
in contrast to Yuan et al. (1998) and Bechtold et al. (2003) studies who
found
a correlation of $\alpha_{\rm ox}$ with redshift.
We detect 32 X-ray emitting galaxies with no sign of AGN activity in
their optical spectra.
They are found to be located in a well defined part of the IR-color-color
diagram. It would be very interesting
to perform more focussed observations to investigate the true nature of
what seems to be
a particular population of X-ray sources.

\section*{Acknowledgments}
We thank the anonymous referee for his/her critical comments that led to 
significant improvement of the paper.  We also thank the present and former staff of the Anglo-Australian Observatory for their work in building
and operationg the AAOmega facility. This work used the CFHTLS data products, which is based on observations
obtained with MegaPrime/MegaCam, a joint project of CFHT and CEA/DAPNIA, at the 
Canada-France-Hawaii Telescope (CFHT) which is operated by the National Research Council (NRC) of Canada,
the Institut National des Science de l'Univers of the Centre National de la Recherche Scientifique (CNRS)
of France, and the Univerisity of Hawaii. This work is based in part on data products produced at TERAPIX
and the Canadian Astronomy Data Centre as part of the Canada-France-Hawaii Telescope Legacy Survey, a collaborative
project of NRC and CNRS.  CSS, PPJ and RS gratefully acknowledge
support from the Indo-French Center for the Promotion of Advanced Research
(Centre Franco-Indien pour la Promotion de la Recherche Avancée) under 
contract No. 3004-3.

\end{document}